\documentclass[aps,prd,twocolumn,nofootinbib,superscriptaddress]{revtex4-1}
\usepackage{graphicx}
\usepackage{color}
\usepackage{natbib}
\usepackage{url}
\usepackage[colorlinks=true]{hyperref}
\usepackage{acronym}
\usepackage[utf8]{inputenc}

\def\msun{M_{\odot}}

\def\flow{f_{\rm low}}

\def\neutronstars{NSs}
\def\blackhole{BH}
\def\blackholes{BHs}

\begin{document}

\title{Designing a template bank to observe compact binary coalescences
in Advanced LIGO's second observing run}

\author{Tito \surname{Dal Canton}}
\email{tito.canton@ligo.org}
\affiliation{Max Planck Institute for Gravitational Physics
             (Albert Einstein Institute), Callinstrasse 38,
             D-30167 Hannover, Germany}
\affiliation{NASA Postdoctoral Program Fellow, Goddard Space Flight Center,
             Greenbelt, MD 20771, USA}
\author{Ian W. Harry}
\email{ian.harry@ligo.org}
\affiliation{Max Planck Institute for Gravitational Physics
			 (Albert Einstein Institute), Am M\"uhlenberg 1,
			 D-14476 Potsdam-Golm, Germany}

\date{\today}

\begin{abstract}
We describe the methodology and novel techniques used to construct a
set of waveforms, or template bank, applicable to searches for compact binary
coalescences in Advanced LIGO's second observing run. This template bank
is suitable for observing systems composed of two neutron stars, two
black holes, or a neutron star and a black hole. The Post-Newtonian
formulation is used to model waveforms with total mass less than 4 $M_{\odot}$ and the
most recent effective-one-body model, calibrated to numerical relativity to include
the merger and ringdown, is used for total masses greater than 4 $M_{\odot}$.
The effects of spin precession, matter, orbital eccentricity and radiation modes
beyond the quadrupole are neglected.
In contrast to the template bank used to search for compact binary mergers in
Advanced LIGO's first observing run, here we are including 
binary-black-hole systems with total mass up to several hundreds of
solar masses, thereby improving the ability to observe such systems. We introduce
a technique to vary the starting frequency of waveform filters so that our bank
can simultaneously contain binary-neutron-star and high-mass binary-black hole waveforms.
We also introduce a lower-bound on the filter waveform length, to exclude very
short-duration, high-mass templates whose sensitivity is
strongly reduced by the characteristics and performance of the interferometers.
\end{abstract}

\maketitle

\acrodef{GW}{gravitational wave}
\acrodef{BH}{black hole}
\acrodef{NS}{neutron star}
\acrodef{NSBH}{neutron-star--black-hole}
\acrodef{SNR}{signal-to-noise ratio}
\acrodef{pN}{post-Newtonian}

\section{Introduction}
\label{sec:intro}
Merging binary systems composed of stellar-mass compact objects are a primary
source of gravitational-wave (GW) signals in ground-based interferometers such as
Advanced LIGO \cite{Harry:2010zz,TheLIGOScientific:2014jea}, Advanced Virgo
\cite{TheVirgo:2014hva} and Kagra \cite{Somiya:2011np,Aso:2013eba}.
The first observing run of Advanced LIGO (O1) yielded two binary-black-hole
(BBH) observations with $> 5\sigma$ signficance, GW150914 and GW151226 and one
further observation with $\sim 2 \sigma$ significance,  LVT151012
\cite{Abbott:2016blz,Abbott:2016nmj,O1BBH}. Many additional observations are
expected in the coming years~\cite{O1BBH}.

The most sensitive search for compact binary mergers leverages accurate physical
models describing the dynamics of such systems and their associated GW signals.
The search parameter space, typically composed of the masses and spin parameters
of the components of the binary, is covered by a discrete grid of
model waveforms called a \emph{template bank} \cite{Sathyaprakash:1991mt,PhysRevD.53.6749,PhysRevD.60.022002,
PhysRevD.76.102004, Prix2007, PhysRevD.80.104014, PhysRevD.89.084041}. Detector data are then
correlated (or matched-filtered) with each waveform in the bank, producing a list
of \emph{triggers}, which are defined by local maxima of the matched-filter
\ac{SNR} being larger than a certain threshold \cite{findchirp}. Each trigger is
tagged with a variety of parameters including the originating template, the time
of merger and a single-detector ranking statistic based on the \ac{SNR}.
Triggers in coincidence between different
detectors are further ranked according to a network statistic and finally
assigned a statistical significance based on empirical background distributions.
Three independently implemented pipelines are currently using such methods to
search for compact binary mergers in Advanced LIGO data: \texttt{PyCBC}
\cite{PhysRevD.90.082004, Usman:2015kfa, alex_nitz_2017_545845, PyCBCPhaseTimeAmpStat},
\texttt{GstLAL} \cite{Cannon:2011vi, Privitera:2013xza, Messick:2016aqy} and
\texttt{MBTA}~\cite{Adams:2015ulm}.

A number of variables are used in the construction of a template bank: (i) the parameter space to
search over (range of masses and spin parameters of the merging objects), (ii)
the power spectral density of the detector noise, (iii) the particular waveform
model(s) to use, (iv) the maximum fractional loss in sensitive
range due to the discrete coverage of the search space with a finite number of
templates and (v) the starting frequency of template waveforms.
These choices impact both the sensitivity of the search as well as its
computational cost~\cite{Keppel:2013yia}.

The all-sky, blind, template-based search for stellar-mass compact binaries in
O1 employed a template bank with a maximum total mass of $100 \msun$
\cite{O1BBH,Abbott:2016ymx}. A second all-sky blind
search targeted BBH systems with larger masses ($50 \msun$ to $600 \msun$) and
employed independently-constructed template banks for the \texttt{gstlal} and
\texttt{PyCBC} analyses \cite{O1IMBH}. The basic search technology employed for
the two searches, however, remained the same. In Advanced LIGO's second
observing run (O2), it is desired to combine these two parameter spaces together
into a single search. In fact, separate and partially-overlapping searches
require artificial boundaries across the continuous parameter space of compact
binary systems. Calculating the statistical significance of candidate events is
also more complicated, as it requires accounting for the same event being
detected by multiple searches as well as accounting for the relative
sensitivities of the involved searches \cite{Biswas:2012ty}. Therefore,
covering the largest
possible parameter space in a single search is more straightforward. Here we describe
a template bank resulting from this decision, whose construction involves a
number of novel  developments. Such a bank is currently being used by the
\texttt{PyCBC} pipeline to search for compact binary mergers in Advanced LIGO's
O2 data.

The paper is organized as follows. Section \ref{sec:construction} describes our
choice of model for the template waveforms, the extent of the search parameter
space, our criterion for defining a variable template starting frequency,
the placement of templates
and the characteristics of the resulting bank. In section \ref{sec:validation}
we demonstrate the ability of the bank to recover signals. We summarize our
conclusions in section \ref{sec:conclusion}. Throughout the paper, the heavier
and lighter objects of a compact binary have masses $m_1$ and $m_2$ and
dimensionless spin components along the orbital angular momentum $\chi_1$ and
$\chi_2$. The binary's total mass and mass ratio are referred to as
$M = m_1 + m_2$ and $q = m_1 / m_2$. Masses are taken to be in the detector
frame, i.e.~redshifted.

\section{Design and construction}
\label{sec:construction}

\subsection{Noise spectral density}

\begin{figure}
  \includegraphics[width=\columnwidth]{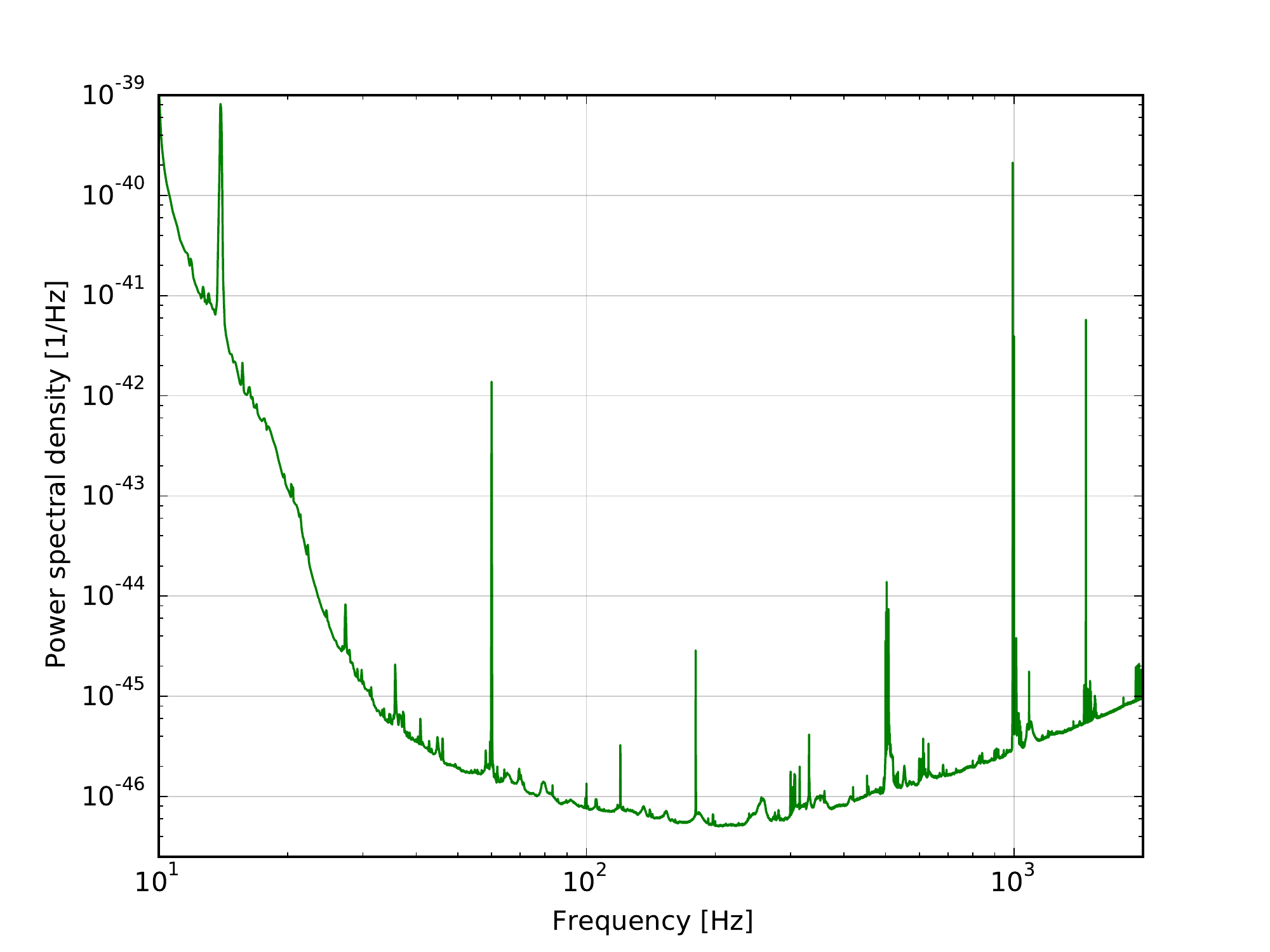}
  \caption{Power spectral density used to model the detector noise in the
  construction of the bank. The curve is the harmonic mean of the power spectral
  densities measured at the Hanford and Livingston detectors.}
  \label{fig:noisecurve}
\end{figure}

The noise power spectral density used for constructing the bank (figure
\ref{fig:noisecurve}) is calculated as the harmonic mean of the power spectral
densities from the Hanford and Livingston detectors, following
\cite{KeppelPlacement}. Averaged single-detector power-spectral densities are estimated from a week
of engineering data acquired immediately before the start of Advanced LIGO's
second observing run following methods described in~\cite{findchirp,Usman:2015kfa}.

\subsection{Waveform models}

The template bank used to analyse O1 data used a reduced-order model~\cite{Purrer:2014fza}
of the aligned-spin effective-one-body waveform model
described in~\cite{Taracchini:2013rva} for templates with total mass $M \ge 4\msun$.
For $M < 4\msun$ the merger
and ringdown portions of the waveform contain a negligible amount of power
relative to the inspiral, so a frequency-domain post-Newtonian inspiral-only model
was used instead, due to simplicity and reduced computational
cost~\cite{Buonanno:2009zt}. The post-Newtonian model included orbital terms up
to 3.5PN order~\cite{Buonanno:2009zt}, and spin-related corrections up to 2.5PN
order~\cite{Arun:2008kb}. It imposed a termination of the waveform at the
frequency of the innermost stable circular orbit of two Schwarzschild \acp{BH}
with total mass $M$.

\begin{figure}
  \includegraphics[width=\columnwidth]{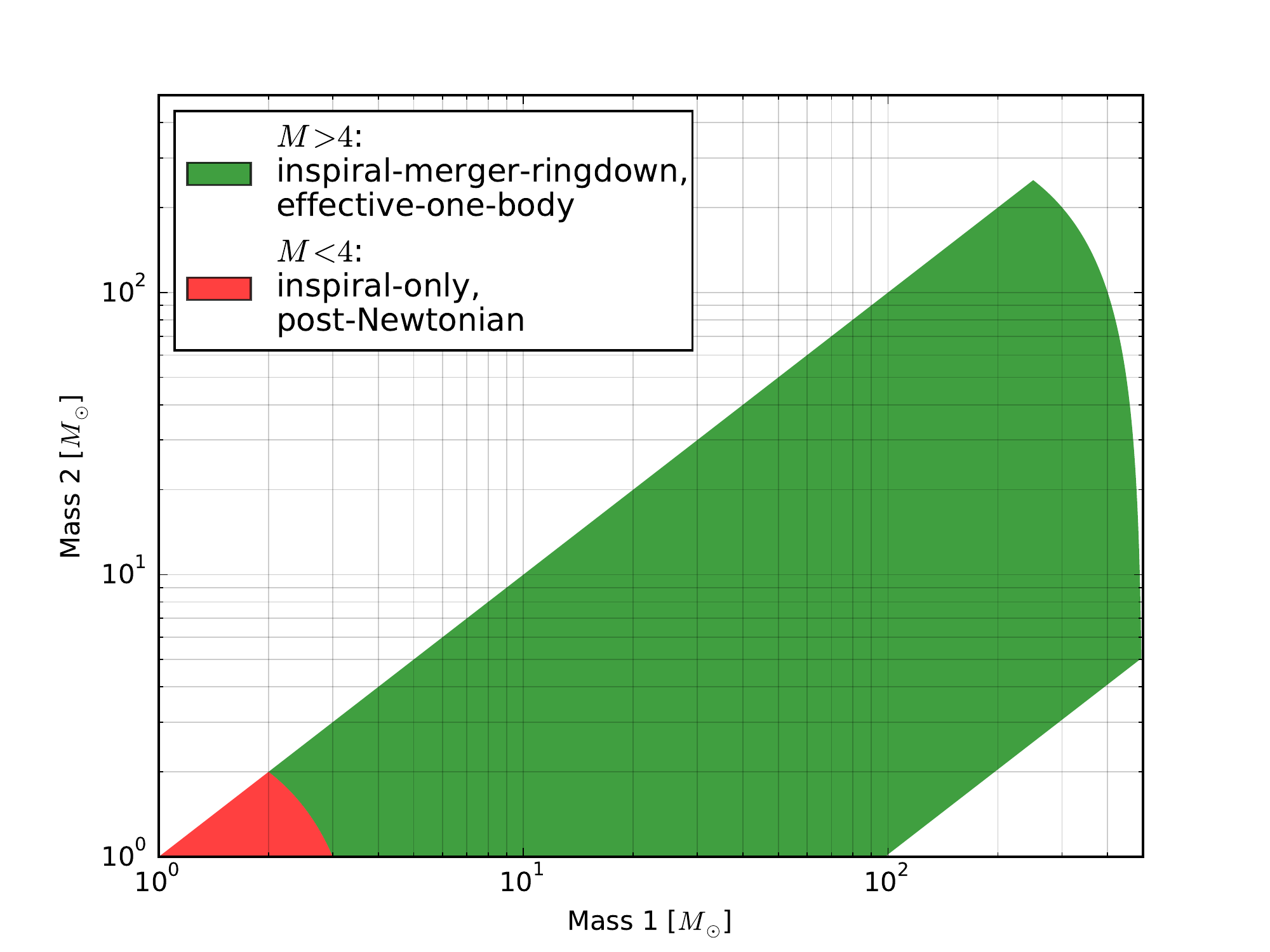}
  \caption{Variation of the template waveform model used to cover the search
  parameter space.}
  \label{fig:approximant}
\end{figure}

Following recent development, for the O2 template
bank a newer version of the effective-one-body waveform model is used~\cite{PhysRevD.95.044028}. The new model is calibrated to a larger set of
numerical-relativity results, which improve it especially for high mass ratios
and high spin magnitudes. We choose to continue using the frequency-domain
post-Newtonian inspiral-only model for $M < 4\msun$, but
the model is updated to include spin effects up to 3.5PN order~\cite{Bohe:2013cla}.
The regions of searched mass space covered by different waveform models are
visualized in figure \ref{fig:approximant}. All waveform
models described in this section, and used in this work, are freely accessible in the
\texttt{lalsimulation} package~\cite{lal}\footnote{
The internal \texttt{lalsimulation} names for the waveform models described here are \texttt{SEOBNRv2}
and \texttt{SEOBNRv2\_ROM\_DoubleSpin} for the effective-one-body model used in the O1 template
bank and its reduced-order representation. \texttt{SEOBNRv4} and \texttt{SEOBNRv4\_ROM} denote
the improved effective-one-body model and reduced-order representation used in the O2 template bank.
\texttt{TaylorF2} is the name of the post-Newtonian waveform model, as described with that name
in \cite{Buonanno:2009zt}.}.

\subsection{Parameter space}

\begin{figure}
  \includegraphics[width=\columnwidth]{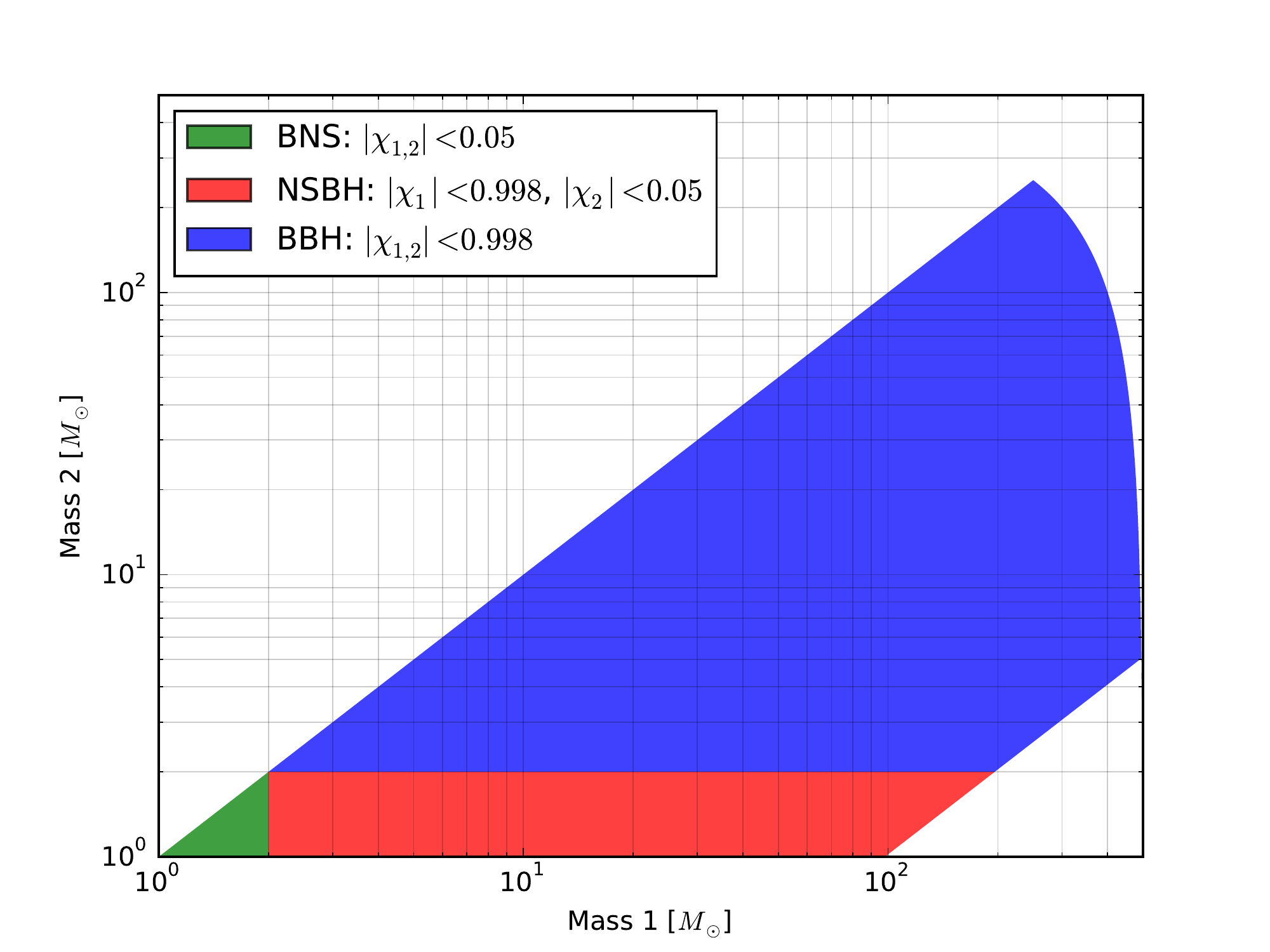}
  \caption{Projection of the O2 search parameter space in the plane of the
  component masses. Colors indicate regions associated with different spin
  limits on the components.}
  \label{fig:searchspace}
\end{figure}

The lower bound on the template bank mass space is a limit on component masses,
specifically that the masses of both components are restricted to be larger
than $1\msun$, matching what was used in O1.
Lowering the component masses even further would rapidly increase the number of
templates, and sub-$\msun$ templates would only be sensitive up to a few tens
of megaparsecs, while there is little expectation of \neutronstars\ with
sub-$\msun$ masses \cite{Miller:2014aaa}.
The mass ratio in O1 was limited to $q \lesssim 98$ as this was the
upper limit for which the effective-one-body waveforms were available. The
same limit exists for the improved model that is used for the O2 template bank.
The high-mass boundary in O1 was simply given as $M < 100\msun$.
In O2 we increase the limit on
total mass to $M < 500 \msun$, therefore including most of the mass range
covered by the O1 search targeting intermediate-mass BBH \cite{O1IMBH}.
As done in O1, components with mass below $2\msun$ are assumed to be
\neutronstars\ and are assigned dimensionless aligned-spin parameters restricted
to spin magnitude $< 0.05$. This enables the search to detect \neutronstars~with
spins up to $0.4$ \cite{AlexNitzThesis} corresponding to the fastest-spinning known
pulsar \cite{Lorimer:2008se}. Heavier components, likely to be \blackholes, are
given spins between $\pm 0.998$ to account for the large spins observed in X-ray
binaries \cite{McClintock:2013vwa}. This is slightly larger than the O1 range
($\pm 0.9895$) thanks to the increase of possible spin values in the improved
effective-one-body waveform model. The boundaries of the search space are shown
in figure \ref{fig:searchspace}.

Extending the mass range is not as simple as raising the maximum total mass, however.
Because of the wide range of mass ratios and spins, high-mass waveforms have a
large variety of time-frequency structure. Waveforms associated with large mass
ratios and high aligned spins complete many cycles in the LIGO sensitive
band, last for a relatively long time and coalesce at relatively large
frequency. Those associated with high \emph{antialigned} spins, instead, merge
at or below the lowest sensitive frequency ($\sim 20$ Hz) and complete only a
few cycles in band, producing a much shorter signal. It has been
found empirically that in the \texttt{PyCBC} search technique, it is difficult
to distinguish if triggers produced by such short-duration templates are due to
instrumental artifacts or due to astrophysical GW signals
\cite{TheLIGOScientific:2016qqj, PyCBCPhaseTimeAmpStat}. Given their short time
scales, such signals would also be efficiently detected by generic transient-GW
searches (see e.g. \cite{Klimenko:2015ypf}) and therefore are not a high priority
target compared to other regions of the parameter space.

\begin{figure}
  \includegraphics[width=\columnwidth]{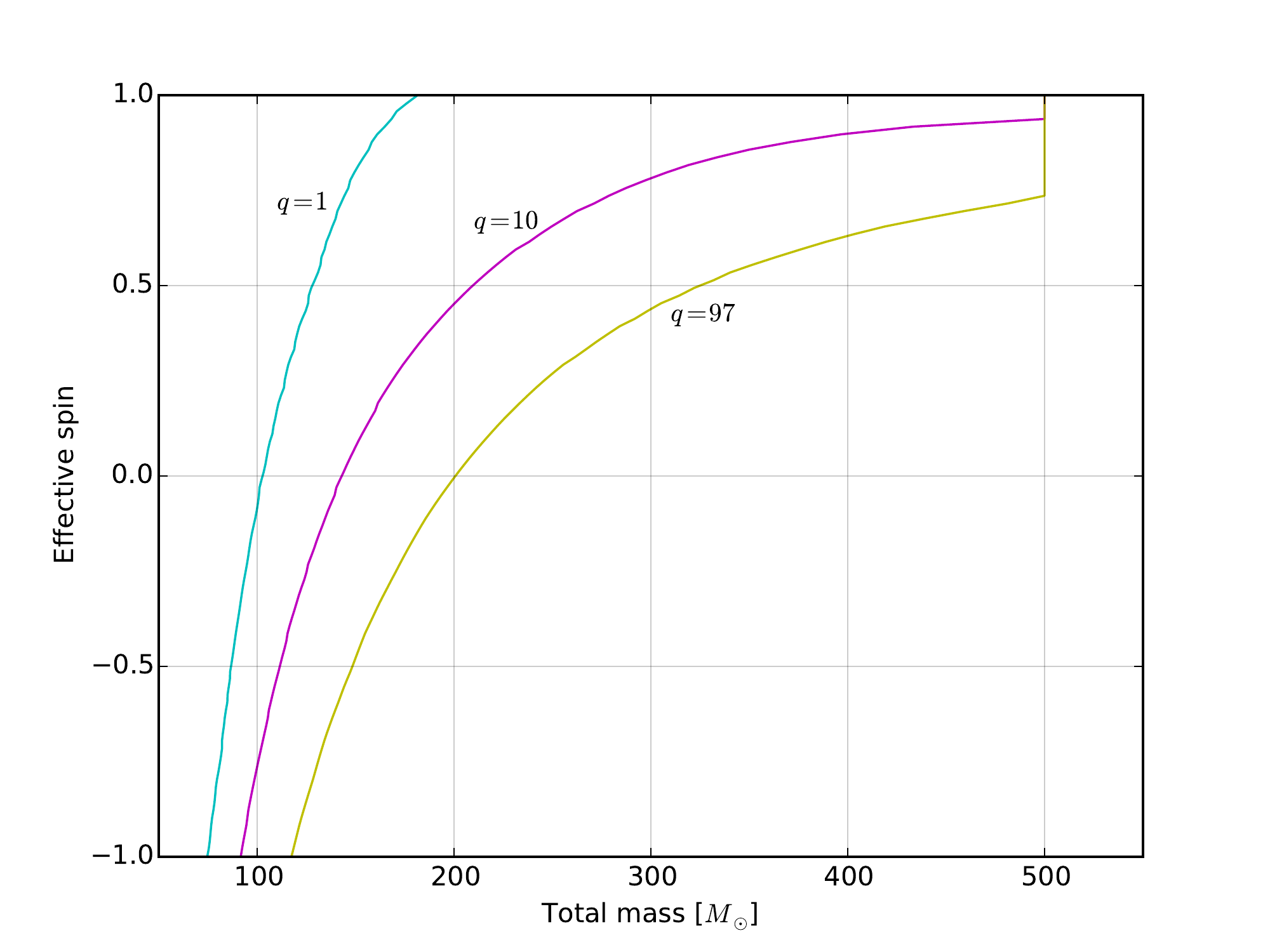}
  \caption{High-mass boundary of the search parameter space as determined by
  the minimum requirement on waveform duration of 150 ms. The boundary is a
  curved surface, spanning a range of total masses, mass ratios and aligned-spin
  components (only the equal-spin case is shown here). The total mass has a
  hard limit at $500\msun$.}
  \label{fig:durationboundary}
\end{figure}

In light of the above considerations we calculate the duration of each waveform
in the time domain and impose a boundary on the parameter space that all waveforms
must have a duration greater than 150 ms. For post-Newtonian templates, the
duration is defined as the usual chirp time \cite{PhysRevD.50.R7111}. For
effective-one-body templates, it is defined as the time taken by the waveform to
go from the starting frequency to the ringdown frequency \cite{Purrer:2015tud}.
The latter definition is increased by $10\%$ in order to account for potentially
long ringdowns. The value of 150 ms is chosen empirically based
on the observed variation of the background distribution in engineering data.
This translates to an irregularly-shaped high-mass boundary for the bank, which
varies with the total mass, the mass ratio and the spin parameters. The maximum
total mass for equal-mass, weakly-spinning templates is $100 \msun$ while
high-mass-ratio, high-aligned-spin templates still have a duration greater than
150 ms at the hard limit of $M=500 \msun$ (figure \ref{fig:durationboundary}).

Further extension of the search space with currently-available waveform
models is challenging. As shown in figure \ref{fig:durationboundary}, increasing
the maximum total mass above $500\msun$ only adds a small extra portion of
parameter space, given the O2 noise curve. The currently available reduced-order
model of the effective-one-body waveforms is also limited to $M \lesssim 500\msun$.
Extending the component dimensionless spins up to the Kerr limit ($\pm 1$) is
possible with the latest effective-one-body waveform model. However, as
described in~\cite{Gralla:2016qfw}, waveforms with
nearly-extremal spins can terminate with quasi-monochromatic features lasting
for many cycles, depending on the mass parameters. Such features in the waveform
greatly reduce the effectiveness of the signal-based consistency tests used
in the \texttt{PyCBC} search method~\cite{chi2veto, PhysRevD.87.024033}. This
effect can potentially reduce the sensitivity of the search in regions of
the parameter space where waveforms have a similar duration to such extreme-spin
templates.
Therefore, extending the \blackhole~spin range all the way to the Kerr limit
requires further investigation of this effect, tests of the reliability of the
waveform models at nearly-extremal spins and potentially the development
of new signal-based consistency tests applicable for such waveforms.

\subsection{Template starting frequency}
\label{sec:flow}

The starting frequency of template waveforms when the bank is constructed
and the lower bound of the matched-filter integral over frequency
\cite{findchirp} are in principle different parameters of the pipeline.
However, we argue that using inconsistent frequencies might reduce either the
effectualness of the bank or the computational efficiency of the search.
Here we assume that both parameters are fixed to the same frequency $\flow$.

The choice of $\flow$ is then a balance of practical considerations.
Reducing $\flow$ has both beneficial and detrimental consequences depending on
which region of the parameter space is being considered. A lower value of
$\flow$ equates to an increase in the sensitivity of the search, assuming Gaussian,
well-calibrated noise, as more of the signal power is recovered by the matched-filter.
The amount by which the sensitivity increases depends on the noise spectral density
and on the template parameters, however.
Below $\sim 20$ Hz the power spectral density of the detector
noise $S(f)$ increases steeply (figure \ref{fig:noisecurve}) and the calibration
of the detectors itself becomes less reliable \cite{PhysRevD.95.062003}. 
When the template terminates at kilohertz frequencies, such as is the case for
binary neutron stars (BNS) and low-mass neutron-star--black-holes (NSBH),
the increase in sensitivity introduced by reducing
$\flow$ will be small. When the merger happens at tens of Hz, such as for BBH
systems with a total mass of several hundred solar masses, the
increase will be larger. Reducing $\flow$ also increases the duration of
template waveforms. Templates lasting hundreds of seconds introduce technical
complications in the implementation of the matched filter, and are more likely
to encounter an instrumental
artifact within their duration, increasing the false-alarm rate or the chance
of contaminating an astrophysical signal \cite{0264-9381-31-1-015016}. For
high-mass BBH waveforms, however, the waveform duration with $\flow = 20$ Hz
will not be more than several seconds, and therefore the chance of intersecting
an instrumental artifact is much lower than for longer BNS waveforms.
Finally, lowering $\flow$ also generally increases the number of required
templates and thus the computational cost of the search, but it will be shown
later that this is not an issue.

Previous matched-filter searches used a fixed $\flow$ for the entire template bank.
For instance, initial-LIGO stellar-mass searches used $\flow = 40$ Hz. Following improvements
in the sensitivity and calibration at low frequency, $\flow$ was reduced to
$30$ Hz for the O1 search described in \cite{O1BBH}, and was set even lower when
searching for BBHs with total masses up to $600 M_{\odot}$ in \cite{O1IMBH}.

The above considerations suggest that wide-mass-range template banks will
benefit from a $\flow$ that varies over the parameter space, being larger for
low-mass waveforms and smaller for higher-mass ones, effectively becoming a
property of each waveform in the template bank.
The criterion we adopt for choosing $\flow$ for each template waveform is the
following:
\begin{equation}
  \frac{R(\flow)}{R(15\ \textrm{Hz})} = 0.995
  \label{eq:flow}
\end{equation}
where
\begin{equation}
  R(f) = \left( \int^{f_{\rm high}}_{f} \frac{|\tilde{h}(\nu)|^2}{S(\nu)} \textrm{d}\nu \right)^{1/2}
\end{equation}
and $\tilde{h}(\nu)$ is the template waveform in the frequency domain and
$f_{\rm high}$ a frequency above which the template has a negligible amount
of observable power (e.g.~the Nyquist frequency). $R(f)$ is
proportional to the sensitive distance of the template calculated using $f$ as the
low-frequency cutoff. Therefore, choosing $\flow$ as in equation (\ref{eq:flow})
means fixing the loss of range ($0.5\%$) with respect to an ideal situation
where data can be analyzed down to 15 Hz.
This criterion is inspired by the traditional choice of determining the minimum
match of the bank based on a fixed loss of range with respect to an infinitely
dense bank.
The reference frequency of 15 Hz is chosen to be lower than the minimum
frequency used so far in a \texttt{PyCBC} analysis
(20 Hz) but still well above the corner frequency of the high-pass filters used
in conditioning the data for estimating the noise spectral density (10 Hz).
Templates which accumulate the bulk of their range at relatively high frequency,
i.e.~BNS and low-mass NSBH,
are automatically assigned a large $\flow$. On the other hand, templates which
coalesce at tens of Hz, such as equal-mass BBH with $M \gtrsim 100\msun$
or unequal-mass BBH with antialigned spins, can only satisfy equation
(\ref{eq:flow}) if $\flow$ is not much larger than the reference frequency of
15 Hz.

\begin{figure}
  \includegraphics[width=\columnwidth]{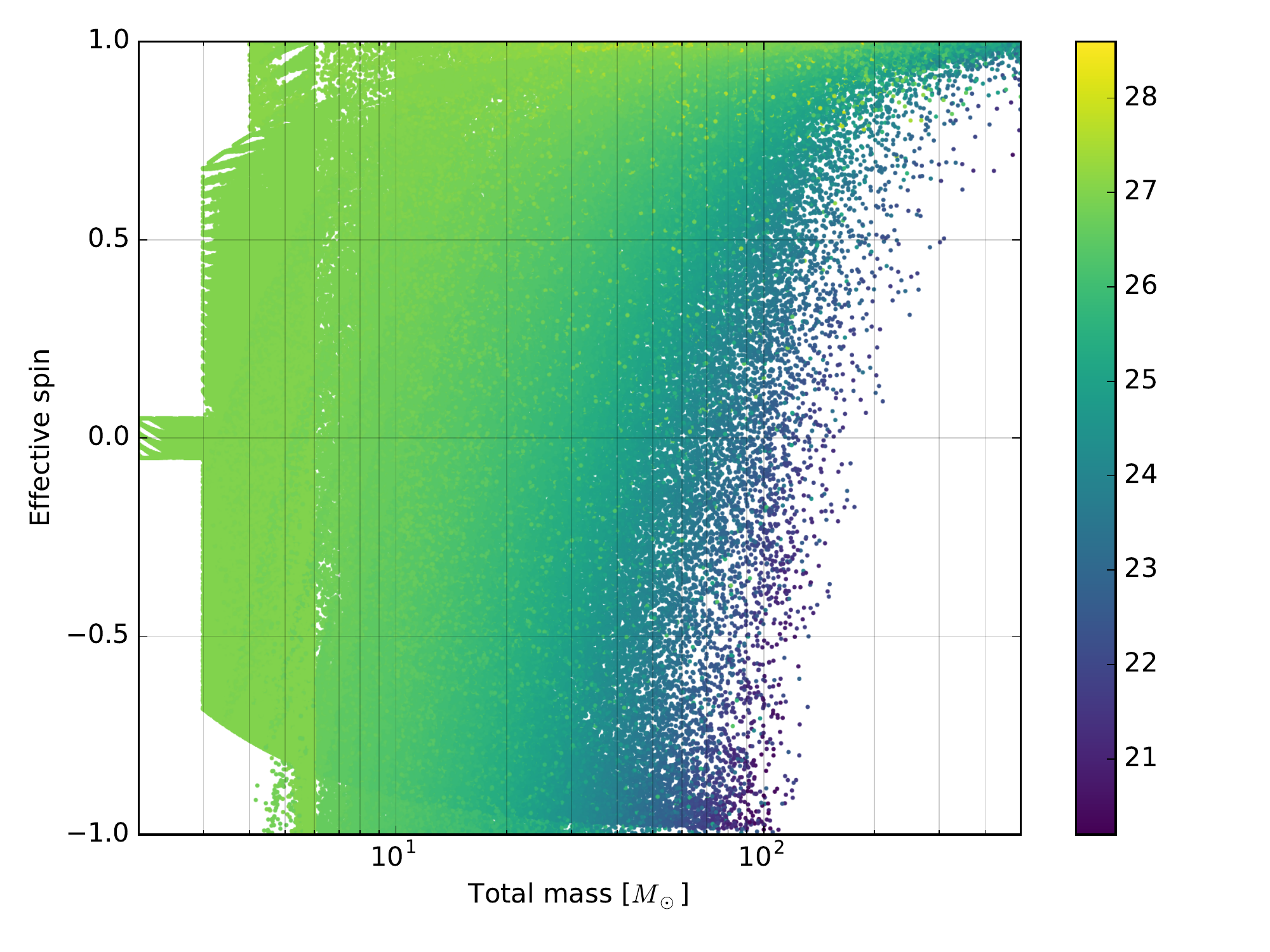}
  \caption{Variation of the low-frequency cutoff assigned to a template
  across the search space. Low-mass templates sweep through the whole sensitive
  band of LIGO and are assigned a lower cutoff at $\sim 27$ Hz. As we go towards
  the high-mass end, templates terminate at progressively lower frequencies and
  their cutoff drops accordingly.}
  \label{fig:varflow}
\end{figure}

The obtained variation of $\flow$ across the total-mass and effective-spin space
is shown in figure \ref{fig:varflow}.
For the lowest-mass templates $\flow = 27$ Hz, close to the fixed value of 30
Hz manually chosen for the O1 bank. The smallest $\flow$ is 20 Hz and
corresponds to the highest-mass templates (at a fixed value of effective spin),
which only have a negligible amount of power at frequencies larger than tens of
Hz.

It is important to note that the variable $\flow$ indirectly affects the
high-mass boundary of the
bank, because the duration of a particular waveform is calculated starting from
the $\flow$ assigned to it. This implies that an improvement in low-frequency
sensitivity, combined with a fixed minimum template duration, automatically
translates to a wider searched mass range. In other words, the search space
naturally adapts to the evolution of the low-frequency performance of the
detectors.

\subsection{Template placement}

\begin{figure}
  \includegraphics[width=\columnwidth]{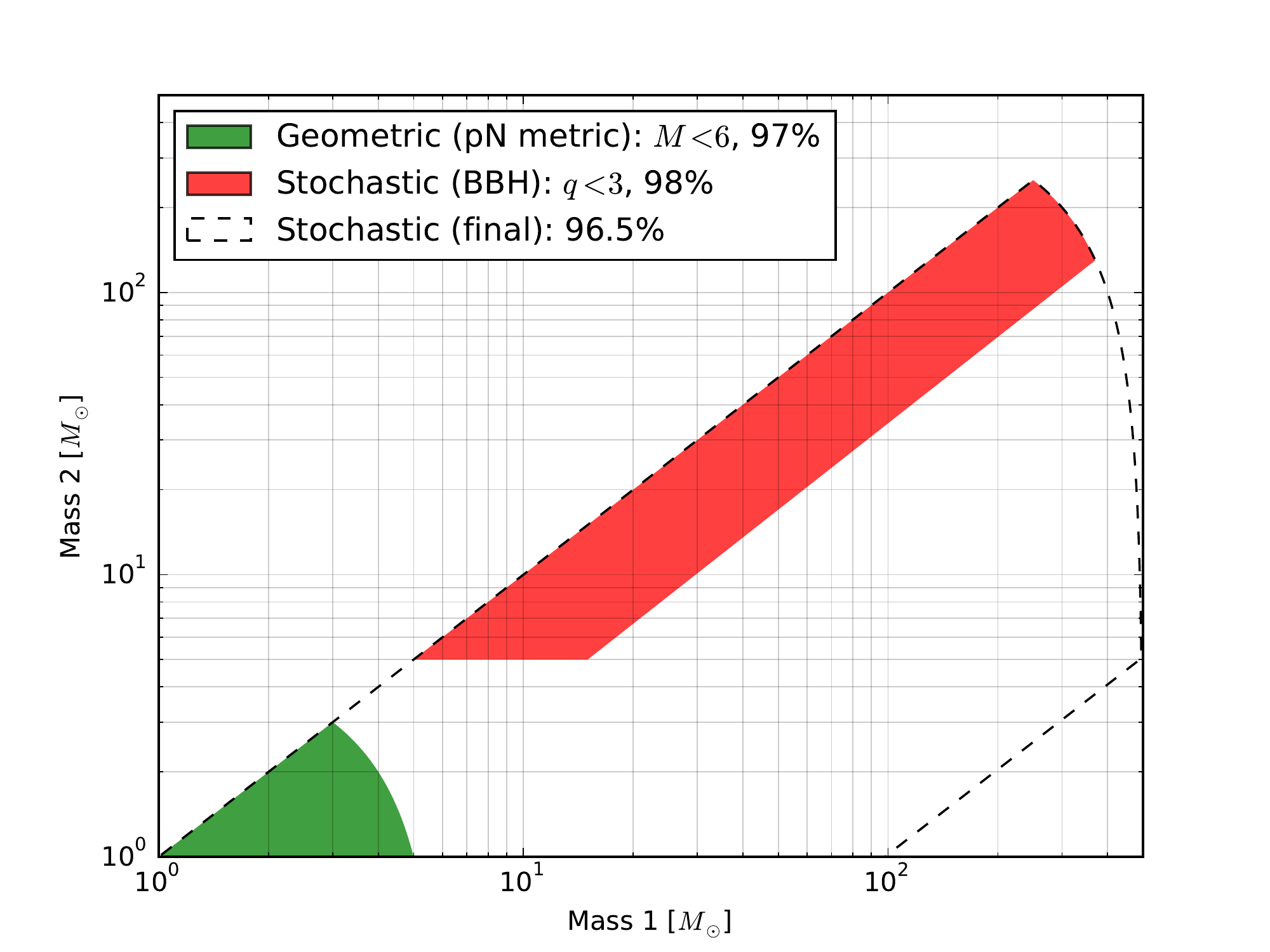}
  \caption{Template placement methods used in covering the full parameter space.}
  \label{fig:placement}
\end{figure}

Templates are placed using the hybrid geometric-stochastic approach described in
\cite{Capano:2016dsf} with some modifications. First, the lowest-mass region of
the search space is covered via geometric placement of post-Newtonian waveform
templates, using a minimum match of $97\%$. At the same time, the high-mass
region with mass ratio less than 3 (i.e.~containing the BBH signals detected
thus far) is filled via stochastic placement using effective-one-body waveform
templates. Since this region has yielded BBH detections before, we place
templates densely here, with a minimum match of $98\%$. The extra coverage
slightly increases the chance of detecting more BBH signals, but it only adds a
small number of templates in comparison to the size of the entire bank. We
combine together these two initial template banks and then ``fill in'' the
remaining parameter space with a third stochastic placement step using
effective-one-body waveform templates and a smaller minimum match of $96.5\%$.
The different placement steps are illustrated in figure \ref{fig:placement}.

One caveat of the initial geometric placement of post-Newtonian templates is
that it must assume fixed frequency limits for computing the parameter-space
metric, so it cannot use a variable $\flow$ as described in section
\ref{sec:flow}. However, in the entire region covered by geometric placement,
equation (\ref{eq:flow}) gives $\flow \approx 27$ Hz to a good approximation.
Therefore we adopt 27 Hz as the $\flow$ of all templates placed geometrically,
as well as for calculating the metric required for geometric placement.
Stochastic placement, on the other hand, simply computes the match between
templates and thus automatically uses whatever $\flow$ is assigned to each
template.

\subsection{Template bank size and structure}

\begin{figure}
  \includegraphics[width=\columnwidth]{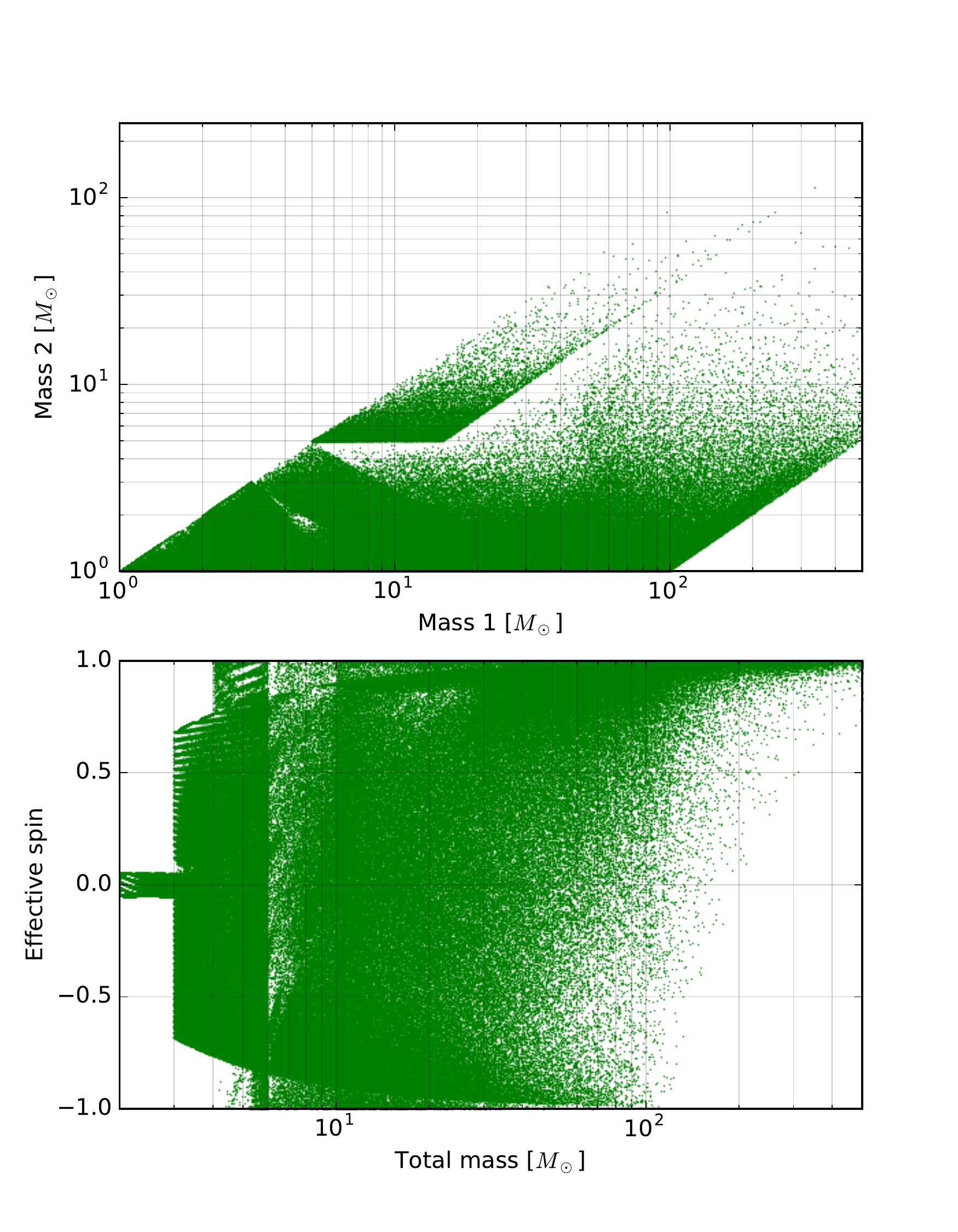}
  \caption{Distribution of templates in two projections of the search space.
  The top panel shows the effect of the different placement steps (compare
  with figure \ref{fig:placement}) while in the lower panel one can see the
  effect of the duration boundary shown in figure \ref{fig:durationboundary}.}
  \label{fig:templatedistr}
\end{figure}

Creating a template bank to cover the parameter space described above and using
the discussed methodology results in $4 \times 10^5$ templates, 60\% larger than
the bank used in O1. The additional computational cost of the pipeline due to
the increased bank size is sufficiently small to not represent a limitation.
The distribution of templates is shown in figure
\ref{fig:templatedistr}. The majority of templates use the effective-one-body,
inspiral-merger-ringdown waveform model and only $7.5 \times 10^4$ templates,
19\% of the bank, are below the $4\msun$ boundary and use post-Newtonian
inspiral-only waveforms. The extra mass space included with respect to the O1
bank ($M>100\msun$) contains $1.7 \times 10^4$ templates, $4.3$\% of the bank.

The ``stripe'' in template distribution at masses larger than $5 \msun$ and
mass ratio between 1 and 3 is the dense initially covered BBH region.
The lack of templates just outside the stripe is due to the fact that templates
at its boundary also cover systems just outside of it, such that the final
covering step does not add templates there.

\section{Validation}
\label{sec:validation}

In a matched-filter search, the maximum \ac{SNR} that can be ideally observed for
a given astrophysical signal (\emph{optimal \ac{SNR}}) requires a template that
exactly matches the signal waveform. Because the template bank is discrete and finite
and because the waveform model does not include the full physics of the system,
the bank can only recover a fraction of the optimal \ac{SNR}, which is known as the
\emph{fitting factor} $\varphi$ between the signal and the bank
\cite{PhysRevD.52.605}. Thus, given a
population of $N$ sources detectable with perfectly-matching templates, only
$\alpha N$ source will be observed on average with a realistic bank, where
$\alpha < 1$ is known as the \emph{signal recovery fraction} \cite{Harry:2016ijz}
and is related to the fitting factor by
\begin{equation}
  \alpha = \frac{\int \mathrm{d}\vec x \ \varphi^3(\vec x) \sigma^3(\vec x)}{\int \mathrm{d}\vec x \ \sigma^3(\vec x)}.
\end{equation}
Here $\vec x$ is the source's parameter vector (excluding the luminosity
distance) and $\sigma(\vec x)$ is the distance at which the optimal \ac{SNR} of
the source takes a fixed reference value\footnote{Normally taken to be 8, in
which case the resulting $\sigma$ is referred to as the \emph{horizon distance}.
However, the choice is arbitrary and does not change the value of the signal
recovery fraction.}. Banks are typically
constructed to achieve at least a $90\%$ signal recovery fraction and it is customary to
evaluate the correct performance of a bank in terms of fitting factor or signal
recovery fraction. This can be done by simulating a large population of compact
binary mergers at a fixed luminosity distance and calculating the fitting factor
between each signal and the bank. Then the signal recovery fraction can be
measured as
\begin{equation}
  \alpha \approx \frac{\sum_i \varphi^3_i \sigma^3_i}{\sum_i \sigma^3_i}
\end{equation}
where $i$ labels each simulated signal.

When testing a bank where all templates use the same lower cutoff frequency,
the optimal \ac{SNR} of each signal is calculated using the same cutoff. As such,
the fitting factor only shows the \ac{SNR} loss due to the discretization of the bank
and any disagreement between the true signal and our waveform model. However, because templates in the
bank described here have a variable $\flow$, optimal \acp{SNR} must now use a
\emph{lower} cutoff, which we choose to be fixed at $15$ Hz, i.e.~the reference
frequency used for calculating each template's $\flow$. Therefore, our fitting
factors also account for the fact that some \ac{SNR} is lost due to a higher starting
frequency of the templates. Since by our definition of $\flow$ this loss is
never smaller than $0.5\%$, our fitting factors cannot be larger than $99.5\%$.

We simulate three different classes of sources: BNS, NSBH and BBH. The BBH set
is split into two subsets by $M=100\msun$, where the lighter set is uniformly
distributed in component masses and covers the BBH mass space used in O1, while
the heavier set is distributed uniformly in $M$ and mass ratio $q$ and covers a
mass range similar to the search space of \cite{O1IMBH}. Each set contains
$5 \times 10^4$ systems and the parameters of the simulations can be found in
table \ref{table:sims}. The waveform model used for the BNS simulations is the
same post-Newtonian model used for templates with $M<4\msun$; NSBH and BBH
simulations use instead the same effective-one-body model used for templates
with $M>4\msun$. Note that the BBH simulations contain signals falling
into the region excluded by the minimum-duration requirement, i.e.~they span a
slightly larger parameter space than the bank is designed to cover.
\begin{table}
  \begin{tabular}{|l|l|l|l|}
  \hline
  Class & Masses ($\msun$) & Aligned spins & Waveform model \\
  \hline
  \hline
  BNS & $m_{1,2} \in [1,3]$ & $\chi_{1,2} \in [\pm 0.05]$ & Post-Newtonian \\
      &                     &                             & \\
  \hline
  NSBH & $m_1 \in [2,50]$ & $\chi_1 \in [\pm 1]$    & Effective-one-body \\
       & $m_2 \in [1,3]$  & $\chi_2 \in [\pm 0.05]$ & \\
  \hline
  BBH & $m_{1,2} > 3$ & $\chi_{1,2} \in [\pm 1]$ & Effective-one-body \\
      & $M < 100$     &                          & \\
  \hline
  BBH & $M \in [100,500]$ & $\chi_{1,2} \in [\pm 1]$ & Effective-one-body \\
      & $q \in [1,10]$    &            & \\
  \hline
  \end{tabular}
  \caption{Parameters of the simulated populations of compact binary mergers
  for testing the bank.}
  \label{table:sims}
\end{table}

\begin{figure*}
  \includegraphics[width=2\columnwidth]{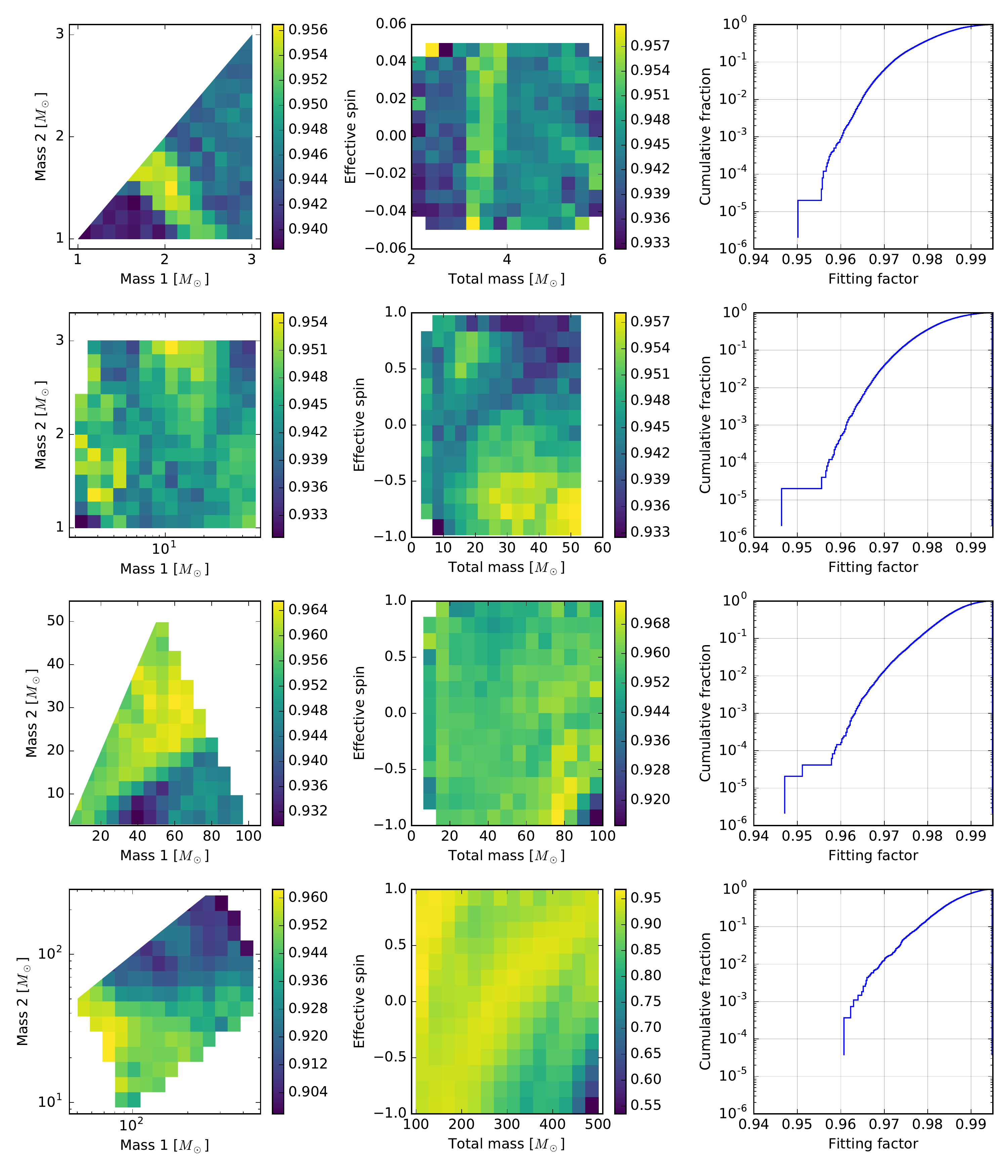}
  \caption{Ability of the bank at detecting signals from BNS (top), NSBH
  (second row), BBH with $M<100\msun$ (third row), and BBH with $M>100\msun$
  (bottom). The heat maps show the signal recovery fraction in each patch of
  parameter space. The cumulative distributions on the right
  show the fitting factors for signals strictly falling within the boundary of
  the search space. Signal recovery fractions are always larger than
  $90\%$, except for negatively-spinning, high-mass BBH systems, which the
  bank does not target. Only $\sim 1\%$ of the signals have an \ac{SNR}
  loss larger than $3.5\%$. The loss is never smaller than $0.5\%$ due to the
  choice of lower cutoff frequency of the templates.}
  \label{fig:banksims}
\end{figure*}

Figure \ref{fig:banksims} presents the signal recovery fractions and fitting
factor distributions for each class.
We divide the component-mass plane into bins and show the signal recovery
fraction in each bin, to highlight possible systematic variations of the
performance of the bank across the search space. The recovery fraction is larger
than 90\% in most bins, with the exception of BBH systems with
$M \gtrsim 200\msun$ and negative effective spins. Those cases, in fact, fall
into the region excluded by the requirement on template duration. If we apply
the same duration cut to exclude the simulated systems in that region, the
resulting fitting factors are larger than $96.5\%$ for 99\% of the systems.
Therefore, the bank meets the design expectations. It is also capable of observing
a large fraction of systems immediately above its high-mass limit.

\section{Conclusion}
\label{sec:conclusion}

In this work we present the motivation, novel methodology and resulting
distribution of the template bank, which is currently being used by the
\texttt{PyCBC} search for compact binary coalescences in Advanced LIGO's O2
data. The method described here creates a template bank covering a mass range
that is as large as practically feasible with current waveform models, search
techniques and detector performance.

In the future, it will be important to include even higher-mass BBH waveforms,
improving the sensitivity to BBH systems heavier than the stellar-mass range and
thus probing potentially different formation mechanisms than those producing
stellar-mass BHs \cite{Miller:2003sc}. As the low-frequency sensitivity of
Advanced LIGO proceeds towards the design target in the coming years, the
duration boundary of the bank proposed here will automatically extend to include
larger masses. It would also be useful to explore new
search techniques for distinguishing more readily between noise transients and
very short-duration compact binary merger signals, therefore lowering the
minimum duration of the templates and raising the maximum mass even further.
Highly-spinning effective-one-body waveforms were also found to be difficult
to distinguish from certain noise artifacts and will require new methods before
the maximum spin magnitude of the current bank can be safely brought to the
Kerr limit.

There are also a number of physical effects that are currently ignored when
creating this template bank. One important approximation is that the component
spins are both restricted to be aligned with the direction of the orbital
angular momentum. This scenario might be preferred if the compact binary is
formed from an isolated stellar binary, but it is also possible that supernovae
kicks can significantly misalign the component spins \cite{Kalogera:1999tq,
Gerosa:2013laa, 0004-637X-742-2-81}.
The compact bodies might also have formed separately in a dense environment,
such as a globular cluster, and later formed a binary with randomly-oriented
spins \cite{Rodriguez:2016vmx}.
The impact of neglecting misaligned spins has been explored
\cite{blo, PhysRevD.91.062010, TheLIGOScientific:2016qqj}, and ideas for
incorporating the effects of precessing spins have been suggested
\cite{Harry:2016ijz, Indik:2016qky}.
Another current approximation is that radiation modes beyond the quadrupole are
neglected in the search templates, and mostly impacts the sensitivity to
asymmetric and high-mass compact binary mergers~\cite{Varma:2014jxa, Bustillo:2016gid}.
Efforts are ongoing to extend the process described here and include these
higher-order effects \cite{Capano:2013raa}. Orbital eccentricity, which presumably
could be found in binaries formed in dense environments, is also currently
neglected. While eccentric waveform models have been recently introduced
\cite{Moore:2016qxz, Tanay:2016zog, Huerta:2016rwp}, a full template bank has
yet to be presented. Several waveform models also incorporate effects due to
NS matter and its equation of state \cite{PhysRevD.83.124008, PhysRevD.88.084011,
PhysRevD.89.043009, Pannarale:2015jka}. Whether and how to extend the current
bank with such waveforms are currently open questions.

\begin{acknowledgments}

We thank Alex Nitz, Tom Dent, Steve Privitera, Alex Nielsen, Alan Weinstein and
Greg Mendell for useful discussion and comments. We are also grateful to
Sarah Caudill for reading through the manuscript and providing thoughtful and
helpful suggestions.

The authors thank the LIGO Scientific Collaboration for access to the data and
gratefully acknowledge the support of the United States National Science
Foundation (NSF) for the construction and operation of the LIGO Laboratory and
Advanced LIGO as well as the Science and Technology Facilities Council (STFC)
of the United Kingdom, and the Max-Planck-Society (MPS) for support of the
construction of Advanced LIGO. Additional support for Advanced LIGO was
provided by the Australian Research Council.

TDC was supported by an appointment to the NASA Postdoctoral Program at the
Goddard Space Flight Center, administered by Universities Space Research
Association under contract with NASA, during part of this work. IWH thanks
the Max Planck Society for support.

Numerical computations and plots in this paper were performed with Python,
NumPy \cite{numpy} and Matplotlib \cite{matplotlib} on the Atlas and Vulcan
clusters of the Max Planck Institute for Gravitational Physics.

This paper has LIGO document number LIGO-P1700085.
\end{acknowledgments}

\bibliography{references}{}

\begin{thebibliography}{72}%
\makeatletter
\providecommand \@ifxundefined [1]{%
 \@ifx{#1\undefined}
}%
\providecommand \@ifnum [1]{%
 \ifnum #1\expandafter \@firstoftwo
 \else \expandafter \@secondoftwo
 \fi
}%
\providecommand \@ifx [1]{%
 \ifx #1\expandafter \@firstoftwo
 \else \expandafter \@secondoftwo
 \fi
}%
\providecommand \natexlab [1]{#1}%
\providecommand \enquote  [1]{``#1''}%
\providecommand \bibnamefont  [1]{#1}%
\providecommand \bibfnamefont [1]{#1}%
\providecommand \citenamefont [1]{#1}%
\providecommand \href@noop [0]{\@secondoftwo}%
\providecommand \href [0]{\begingroup \@sanitize@url \@href}%
\providecommand \@href[1]{\@@startlink{#1}\@@href}%
\providecommand \@@href[1]{\endgroup#1\@@endlink}%
\providecommand \@sanitize@url [0]{\catcode `\\12\catcode `\$12\catcode
  `\&12\catcode `\#12\catcode `\^12\catcode `\_12\catcode `\%12\relax}%
\providecommand \@@startlink[1]{}%
\providecommand \@@endlink[0]{}%
\providecommand \url  [0]{\begingroup\@sanitize@url \@url }%
\providecommand \@url [1]{\endgroup\@href {#1}{\urlprefix }}%
\providecommand \urlprefix  [0]{URL }%
\providecommand \Eprint [0]{\href }%
\providecommand \doibase [0]{http://dx.doi.org/}%
\providecommand \selectlanguage [0]{\@gobble}%
\providecommand \bibinfo  [0]{\@secondoftwo}%
\providecommand \bibfield  [0]{\@secondoftwo}%
\providecommand \translation [1]{[#1]}%
\providecommand \BibitemOpen [0]{}%
\providecommand \bibitemStop [0]{}%
\providecommand \bibitemNoStop [0]{.\EOS\space}%
\providecommand \EOS [0]{\spacefactor3000\relax}%
\providecommand \BibitemShut  [1]{\csname bibitem#1\endcsname}%
\let\auto@bib@innerbib\@empty
%</preamble>
\bibitem [{\citenamefont {Harry}(2010)}]{Harry:2010zz}%
  \BibitemOpen
  \bibfield  {author} {\bibinfo {author} {\bibfnamefont {G.~M.}\ \bibnamefont
  {Harry}} (\bibinfo {collaboration} {LIGO Scientific}),\ }\bibfield
  {booktitle} {\emph {\bibinfo {booktitle} {{Gravitational waves. Proceedings,
  8th Edoardo Amaldi Conference, Amaldi 8, New York, USA, June 22-26, 2009}}},\
  }\href {\doibase 10.1088/0264-9381/27/8/084006} {\bibfield  {journal}
  {\bibinfo  {journal} {Class. Quant. Grav.}\ }\textbf {\bibinfo {volume}
  {27}},\ \bibinfo {pages} {084006} (\bibinfo {year} {2010})}\BibitemShut
  {NoStop}%
%%CITATION = CQGRD,27,084006;%%
\bibitem [{\citenamefont {Aasi}\ \emph {et~al.}(2015)\citenamefont {Aasi} \emph
  {et~al.}}]{TheLIGOScientific:2014jea}%
  \BibitemOpen
  \bibfield  {author} {\bibinfo {author} {\bibfnamefont {J.}~\bibnamefont
  {Aasi}} \emph {et~al.} (\bibinfo {collaboration} {LIGO Scientific}),\ }\href
  {\doibase 10.1088/0264-9381/32/7/074001} {\bibfield  {journal} {\bibinfo
  {journal} {Class. Quant. Grav.}\ }\textbf {\bibinfo {volume} {32}},\ \bibinfo
  {pages} {074001} (\bibinfo {year} {2015})},\ \Eprint
  {http://arxiv.org/abs/1411.4547} {arXiv:1411.4547 [gr-qc]} \BibitemShut
  {NoStop}%
%%CITATION = ARXIV:1411.4547;%%
\bibitem [{\citenamefont {Acernese}\ \emph {et~al.}(2015)\citenamefont
  {Acernese} \emph {et~al.}}]{TheVirgo:2014hva}%
  \BibitemOpen
  \bibfield  {author} {\bibinfo {author} {\bibfnamefont {F.}~\bibnamefont
  {Acernese}} \emph {et~al.} (\bibinfo {collaboration} {VIRGO}),\ }\href
  {\doibase 10.1088/0264-9381/32/2/024001} {\bibfield  {journal} {\bibinfo
  {journal} {Class. Quant. Grav.}\ }\textbf {\bibinfo {volume} {32}},\ \bibinfo
  {pages} {024001} (\bibinfo {year} {2015})},\ \Eprint
  {http://arxiv.org/abs/1408.3978} {arXiv:1408.3978 [gr-qc]} \BibitemShut
  {NoStop}%
%%CITATION = ARXIV:1408.3978;%%
\bibitem [{\citenamefont {Somiya}(2012)}]{Somiya:2011np}%
  \BibitemOpen
  \bibfield  {author} {\bibinfo {author} {\bibfnamefont {K.}~\bibnamefont
  {Somiya}} (\bibinfo {collaboration} {KAGRA}),\ }\bibfield  {booktitle} {\emph
  {\bibinfo {booktitle} {{Gravitational waves. Numerical relativity - data
  analysis. Proceedings, 9th Edoardo Amaldi Conference, Amaldi 9, and meeting,
  NRDA 2011, Cardiff, UK, July 10-15, 2011}}},\ }\href {\doibase
  10.1088/0264-9381/29/12/124007} {\bibfield  {journal} {\bibinfo  {journal}
  {Class. Quant. Grav.}\ }\textbf {\bibinfo {volume} {29}},\ \bibinfo {pages}
  {124007} (\bibinfo {year} {2012})},\ \Eprint {http://arxiv.org/abs/1111.7185}
  {arXiv:1111.7185 [gr-qc]} \BibitemShut {NoStop}%
%%CITATION = ARXIV:1111.7185;%%
\bibitem [{\citenamefont {Aso}\ \emph {et~al.}(2013)\citenamefont {Aso},
  \citenamefont {Michimura}, \citenamefont {Somiya}, \citenamefont {Ando},
  \citenamefont {Miyakawa}, \citenamefont {Sekiguchi}, \citenamefont
  {Tatsumi},\ and\ \citenamefont {Yamamoto}}]{Aso:2013eba}%
  \BibitemOpen
  \bibfield  {author} {\bibinfo {author} {\bibfnamefont {Y.}~\bibnamefont
  {Aso}}, \bibinfo {author} {\bibfnamefont {Y.}~\bibnamefont {Michimura}},
  \bibinfo {author} {\bibfnamefont {K.}~\bibnamefont {Somiya}}, \bibinfo
  {author} {\bibfnamefont {M.}~\bibnamefont {Ando}}, \bibinfo {author}
  {\bibfnamefont {O.}~\bibnamefont {Miyakawa}}, \bibinfo {author}
  {\bibfnamefont {T.}~\bibnamefont {Sekiguchi}}, \bibinfo {author}
  {\bibfnamefont {D.}~\bibnamefont {Tatsumi}}, \ and\ \bibinfo {author}
  {\bibfnamefont {H.}~\bibnamefont {Yamamoto}} (\bibinfo {collaboration}
  {KAGRA}),\ }\href {\doibase 10.1103/PhysRevD.88.043007} {\bibfield  {journal}
  {\bibinfo  {journal} {Phys. Rev.}\ }\textbf {\bibinfo {volume} {D88}},\
  \bibinfo {pages} {043007} (\bibinfo {year} {2013})},\ \Eprint
  {http://arxiv.org/abs/1306.6747} {arXiv:1306.6747 [gr-qc]} \BibitemShut
  {NoStop}%
%%CITATION = ARXIV:1306.6747;%%
\bibitem [{\citenamefont {Abbott}\ \emph
  {et~al.}(2016{\natexlab{a}})\citenamefont {Abbott} \emph
  {et~al.}}]{Abbott:2016blz}%
  \BibitemOpen
  \bibfield  {author} {\bibinfo {author} {\bibfnamefont {B.~P.}\ \bibnamefont
  {Abbott}} \emph {et~al.} (\bibinfo {collaboration} {Virgo, LIGO
  Scientific}),\ }\href {\doibase 10.1103/PhysRevLett.116.061102} {\bibfield
  {journal} {\bibinfo  {journal} {Phys. Rev. Lett.}\ }\textbf {\bibinfo
  {volume} {116}},\ \bibinfo {pages} {061102} (\bibinfo {year}
  {2016}{\natexlab{a}})},\ \Eprint {http://arxiv.org/abs/1602.03837}
  {arXiv:1602.03837 [gr-qc]} \BibitemShut {NoStop}%
%%CITATION = ARXIV:1602.03837;%%
\bibitem [{\citenamefont {Abbott}\ \emph
  {et~al.}(2016{\natexlab{b}})\citenamefont {Abbott} \emph
  {et~al.}}]{Abbott:2016nmj}%
  \BibitemOpen
  \bibfield  {author} {\bibinfo {author} {\bibfnamefont {B.~P.}\ \bibnamefont
  {Abbott}} \emph {et~al.} (\bibinfo {collaboration} {Virgo, LIGO
  Scientific}),\ }\href {\doibase 10.1103/PhysRevLett.116.241103} {\bibfield
  {journal} {\bibinfo  {journal} {Phys. Rev. Lett.}\ }\textbf {\bibinfo
  {volume} {116}},\ \bibinfo {pages} {241103} (\bibinfo {year}
  {2016}{\natexlab{b}})},\ \Eprint {http://arxiv.org/abs/1606.04855}
  {arXiv:1606.04855 [gr-qc]} \BibitemShut {NoStop}%
%%CITATION = ARXIV:1606.04855;%%
\bibitem [{\citenamefont {Abbott}\ \emph
  {et~al.}(2016{\natexlab{c}})\citenamefont {Abbott} \emph {et~al.}}]{O1BBH}%
  \BibitemOpen
  \bibfield  {author} {\bibinfo {author} {\bibfnamefont {B.~P.}\ \bibnamefont
  {Abbott}} \emph {et~al.} (\bibinfo {collaboration} {Virgo, LIGO
  Scientific}),\ }\href {\doibase 10.1103/PhysRevX.6.041015} {\bibfield
  {journal} {\bibinfo  {journal} {Phys. Rev. X}\ }\textbf {\bibinfo {volume}
  {6}},\ \bibinfo {pages} {041015} (\bibinfo {year} {2016}{\natexlab{c}})},\
  \Eprint {http://arxiv.org/abs/1606.04856} {arXiv:1606.04856 [gr-qc]}
  \BibitemShut {NoStop}%
%%CITATION = ARXIV:1606.04856;%%
\bibitem [{\citenamefont {Sathyaprakash}\ and\ \citenamefont
  {Dhurandhar}(1991)}]{Sathyaprakash:1991mt}%
  \BibitemOpen
  \bibfield  {author} {\bibinfo {author} {\bibfnamefont {B.~S.}\ \bibnamefont
  {Sathyaprakash}}\ and\ \bibinfo {author} {\bibfnamefont {S.~V.}\ \bibnamefont
  {Dhurandhar}},\ }\href {\doibase 10.1103/PhysRevD.44.3819} {\bibfield
  {journal} {\bibinfo  {journal} {Phys. Rev.}\ }\textbf {\bibinfo {volume}
  {D44}},\ \bibinfo {pages} {3819} (\bibinfo {year} {1991})}\BibitemShut
  {NoStop}%
%%CITATION = PHRVA,D44,3819;%%
\bibitem [{\citenamefont {Owen}(1996)}]{PhysRevD.53.6749}%
  \BibitemOpen
  \bibfield  {author} {\bibinfo {author} {\bibfnamefont {B.~J.}\ \bibnamefont
  {Owen}},\ }\href {\doibase 10.1103/PhysRevD.53.6749} {\bibfield  {journal}
  {\bibinfo  {journal} {Phys. Rev. D}\ }\textbf {\bibinfo {volume} {53}},\
  \bibinfo {pages} {6749} (\bibinfo {year} {1996})}\BibitemShut {NoStop}%
\bibitem [{\citenamefont {Owen}\ and\ \citenamefont
  {Sathyaprakash}(1999)}]{PhysRevD.60.022002}%
  \BibitemOpen
  \bibfield  {author} {\bibinfo {author} {\bibfnamefont {B.~J.}\ \bibnamefont
  {Owen}}\ and\ \bibinfo {author} {\bibfnamefont {B.~S.}\ \bibnamefont
  {Sathyaprakash}},\ }\href {\doibase 10.1103/PhysRevD.60.022002} {\bibfield
  {journal} {\bibinfo  {journal} {Phys. Rev. D}\ }\textbf {\bibinfo {volume}
  {60}},\ \bibinfo {pages} {022002} (\bibinfo {year} {1999})}\BibitemShut
  {NoStop}%
\bibitem [{\citenamefont {Cokelaer}(2007)}]{PhysRevD.76.102004}%
  \BibitemOpen
  \bibfield  {author} {\bibinfo {author} {\bibfnamefont {T.}~\bibnamefont
  {Cokelaer}},\ }\href {\doibase 10.1103/PhysRevD.76.102004} {\bibfield
  {journal} {\bibinfo  {journal} {Phys. Rev. D}\ }\textbf {\bibinfo {volume}
  {76}},\ \bibinfo {pages} {102004} (\bibinfo {year} {2007})}\BibitemShut
  {NoStop}%
\bibitem [{\citenamefont {Prix}(2007)}]{Prix2007}%
  \BibitemOpen
  \bibfield  {author} {\bibinfo {author} {\bibfnamefont {R.}~\bibnamefont
  {Prix}},\ }\href {http://stacks.iop.org/0264-9381/24/i=19/a=S11} {\bibfield
  {journal} {\bibinfo  {journal} {Classical and Quantum Gravity}\ }\textbf
  {\bibinfo {volume} {24}},\ \bibinfo {pages} {S481} (\bibinfo {year}
  {2007})},\ \Eprint {http://arxiv.org/abs/0707.0428} {arXiv:0707.0428 [gr-qc]}
  \BibitemShut {NoStop}%
\bibitem [{\citenamefont {Harry}\ \emph {et~al.}(2009)\citenamefont {Harry},
  \citenamefont {Allen},\ and\ \citenamefont
  {Sathyaprakash}}]{PhysRevD.80.104014}%
  \BibitemOpen
  \bibfield  {author} {\bibinfo {author} {\bibfnamefont {I.~W.}\ \bibnamefont
  {Harry}}, \bibinfo {author} {\bibfnamefont {B.}~\bibnamefont {Allen}}, \ and\
  \bibinfo {author} {\bibfnamefont {B.~S.}\ \bibnamefont {Sathyaprakash}},\
  }\href {\doibase 10.1103/PhysRevD.80.104014} {\bibfield  {journal} {\bibinfo
  {journal} {Phys. Rev. D}\ }\textbf {\bibinfo {volume} {80}},\ \bibinfo
  {pages} {104014} (\bibinfo {year} {2009})}\BibitemShut {NoStop}%
\bibitem [{\citenamefont {Ajith}\ \emph {et~al.}(2014)\citenamefont {Ajith},
  \citenamefont {Fotopoulos}, \citenamefont {Privitera}, \citenamefont
  {Neunzert}, \citenamefont {Mazumder},\ and\ \citenamefont
  {Weinstein}}]{PhysRevD.89.084041}%
  \BibitemOpen
  \bibfield  {author} {\bibinfo {author} {\bibfnamefont {P.}~\bibnamefont
  {Ajith}}, \bibinfo {author} {\bibfnamefont {N.}~\bibnamefont {Fotopoulos}},
  \bibinfo {author} {\bibfnamefont {S.}~\bibnamefont {Privitera}}, \bibinfo
  {author} {\bibfnamefont {A.}~\bibnamefont {Neunzert}}, \bibinfo {author}
  {\bibfnamefont {N.}~\bibnamefont {Mazumder}}, \ and\ \bibinfo {author}
  {\bibfnamefont {A.~J.}\ \bibnamefont {Weinstein}},\ }\href {\doibase
  10.1103/PhysRevD.89.084041} {\bibfield  {journal} {\bibinfo  {journal} {Phys.
  Rev. D}\ }\textbf {\bibinfo {volume} {89}},\ \bibinfo {pages} {084041}
  (\bibinfo {year} {2014})}\BibitemShut {NoStop}%
\bibitem [{\citenamefont {Allen}\ \emph {et~al.}(2012)\citenamefont {Allen},
  \citenamefont {Anderson}, \citenamefont {Brady}, \citenamefont {Brown},\ and\
  \citenamefont {Creighton}}]{findchirp}%
  \BibitemOpen
  \bibfield  {author} {\bibinfo {author} {\bibfnamefont {B.}~\bibnamefont
  {Allen}}, \bibinfo {author} {\bibfnamefont {W.~G.}\ \bibnamefont {Anderson}},
  \bibinfo {author} {\bibfnamefont {P.~R.}\ \bibnamefont {Brady}}, \bibinfo
  {author} {\bibfnamefont {D.~A.}\ \bibnamefont {Brown}}, \ and\ \bibinfo
  {author} {\bibfnamefont {J.~D.~E.}\ \bibnamefont {Creighton}},\ }\href
  {\doibase 10.1103/PhysRevD.85.122006} {\bibfield  {journal} {\bibinfo
  {journal} {Phys. Rev. D}\ }\textbf {\bibinfo {volume} {85}},\ \bibinfo
  {pages} {122006} (\bibinfo {year} {2012})},\ \Eprint
  {http://arxiv.org/abs/gr-qc/0509116} {arXiv:gr-qc/0509116 [gr-qc]}
  \BibitemShut {NoStop}%
%%CITATION = GR-QC/0509116;%%
\bibitem [{\citenamefont {Dal~Canton}\ \emph
  {et~al.}(2014{\natexlab{a}})\citenamefont {Dal~Canton}, \citenamefont {Nitz},
  \citenamefont {Lundgren}, \citenamefont {Nielsen} \emph
  {et~al.}}]{PhysRevD.90.082004}%
  \BibitemOpen
  \bibfield  {author} {\bibinfo {author} {\bibfnamefont {T.}~\bibnamefont
  {Dal~Canton}}, \bibinfo {author} {\bibfnamefont {A.~H.}\ \bibnamefont
  {Nitz}}, \bibinfo {author} {\bibfnamefont {A.~P.}\ \bibnamefont {Lundgren}},
  \bibinfo {author} {\bibfnamefont {A.~B.}\ \bibnamefont {Nielsen}},  \emph
  {et~al.},\ }\href {\doibase 10.1103/PhysRevD.90.082004} {\bibfield  {journal}
  {\bibinfo  {journal} {Phys. Rev. D}\ }\textbf {\bibinfo {volume} {90}},\
  \bibinfo {pages} {082004} (\bibinfo {year} {2014}{\natexlab{a}})},\ \Eprint
  {http://arxiv.org/abs/1405.6731} {arXiv:1405.6731 [gr-qc]} \BibitemShut
  {NoStop}%
\bibitem [{\citenamefont {Usman}\ \emph {et~al.}(2016)\citenamefont {Usman}
  \emph {et~al.}}]{Usman:2015kfa}%
  \BibitemOpen
  \bibfield  {author} {\bibinfo {author} {\bibfnamefont {S.~A.}\ \bibnamefont
  {Usman}} \emph {et~al.},\ }\href {\doibase 10.1088/0264-9381/33/21/215004}
  {\bibfield  {journal} {\bibinfo  {journal} {Class. Quant. Grav.}\ }\textbf
  {\bibinfo {volume} {33}},\ \bibinfo {pages} {215004} (\bibinfo {year}
  {2016})},\ \Eprint {http://arxiv.org/abs/1508.02357} {arXiv:1508.02357
  [gr-qc]} \BibitemShut {NoStop}%
%%CITATION = ARXIV:1508.02357;%%
\bibitem [{\citenamefont {Nitz}\ \emph
  {et~al.}(2017{\natexlab{a}})\citenamefont {Nitz}, \citenamefont {Harry},
  \citenamefont {Biwer}, \citenamefont {Brown}, \citenamefont {Willis},
  \citenamefont {Canton}, \citenamefont {Pekowsky}, \citenamefont {Dent},
  \citenamefont {Williamson}, \citenamefont {Capano}, \citenamefont {De},
  \citenamefont {Kumar}, \citenamefont {Machenschalk}, \citenamefont {Cabero},
  \citenamefont {Massinger}, \citenamefont {Lenon}, \citenamefont {Fairhurst},
  \citenamefont {Reyes}, \citenamefont {Nielsen}, \citenamefont {Kapadia},
  \citenamefont {Pannarale}, \citenamefont {Singer}, \citenamefont {Babak},
  \citenamefont {Macleod}, \citenamefont {Sugar}, \citenamefont {na~Zertuche},
  \citenamefont {Veitch}, \citenamefont {Couvares}, \citenamefont {Bockelman},\
  and\ \citenamefont {Brown}}]{alex_nitz_2017_545845}%
  \BibitemOpen
  \bibfield  {author} {\bibinfo {author} {\bibfnamefont {A.~H.}\ \bibnamefont
  {Nitz}}, \bibinfo {author} {\bibfnamefont {I.}~\bibnamefont {Harry}},
  \bibinfo {author} {\bibfnamefont {C.~M.}\ \bibnamefont {Biwer}}, \bibinfo
  {author} {\bibfnamefont {D.~A.}\ \bibnamefont {Brown}}, \bibinfo {author}
  {\bibfnamefont {J.}~\bibnamefont {Willis}}, \bibinfo {author} {\bibfnamefont
  {T.~D.}\ \bibnamefont {Canton}}, \bibinfo {author} {\bibfnamefont
  {L.}~\bibnamefont {Pekowsky}}, \bibinfo {author} {\bibfnamefont
  {T.}~\bibnamefont {Dent}}, \bibinfo {author} {\bibfnamefont {A.~R.}\
  \bibnamefont {Williamson}}, \bibinfo {author} {\bibfnamefont
  {C.}~\bibnamefont {Capano}}, \bibinfo {author} {\bibfnamefont
  {S.}~\bibnamefont {De}}, \bibinfo {author} {\bibfnamefont {P.}~\bibnamefont
  {Kumar}}, \bibinfo {author} {\bibfnamefont {B.}~\bibnamefont {Machenschalk}},
  \bibinfo {author} {\bibfnamefont {M.}~\bibnamefont {Cabero}}, \bibinfo
  {author} {\bibfnamefont {T.}~\bibnamefont {Massinger}}, \bibinfo {author}
  {\bibfnamefont {A.}~\bibnamefont {Lenon}}, \bibinfo {author} {\bibfnamefont
  {S.}~\bibnamefont {Fairhurst}}, \bibinfo {author} {\bibfnamefont
  {S.}~\bibnamefont {Reyes}}, \bibinfo {author} {\bibfnamefont
  {A.}~\bibnamefont {Nielsen}}, \bibinfo {author} {\bibfnamefont
  {S.}~\bibnamefont {Kapadia}}, \bibinfo {author} {\bibfnamefont
  {F.}~\bibnamefont {Pannarale}}, \bibinfo {author} {\bibfnamefont
  {L.}~\bibnamefont {Singer}}, \bibinfo {author} {\bibfnamefont
  {S.}~\bibnamefont {Babak}}, \bibinfo {author} {\bibfnamefont
  {D.}~\bibnamefont {Macleod}}, \bibinfo {author} {\bibfnamefont
  {C.}~\bibnamefont {Sugar}}, \bibinfo {author} {\bibfnamefont {L.~M.}\
  \bibnamefont {na~Zertuche}}, \bibinfo {author} {\bibfnamefont
  {J.}~\bibnamefont {Veitch}}, \bibinfo {author} {\bibfnamefont
  {P.}~\bibnamefont {Couvares}}, \bibinfo {author} {\bibfnamefont
  {B.}~\bibnamefont {Bockelman}}, \ and\ \bibinfo {author} {\bibfnamefont
  {N.~W.}\ \bibnamefont {Brown}},\ }\href {\doibase 10.5281/zenodo.545845}
  {\enquote {\bibinfo {title} {Pycbc software},}\ } (\bibinfo {year}
  {2017}{\natexlab{a}})\BibitemShut {NoStop}%
\bibitem [{\citenamefont {Nitz}\ \emph
  {et~al.}(2017{\natexlab{b}})\citenamefont {Nitz}, \citenamefont {Dent},
  \citenamefont {Dal~Canton}, \citenamefont {Fairhurst},\ and\ \citenamefont
  {Brown}}]{PyCBCPhaseTimeAmpStat}%
  \BibitemOpen
  \bibfield  {author} {\bibinfo {author} {\bibfnamefont {A.}~\bibnamefont
  {Nitz}}, \bibinfo {author} {\bibfnamefont {T.}~\bibnamefont {Dent}}, \bibinfo
  {author} {\bibfnamefont {T.}~\bibnamefont {Dal~Canton}}, \bibinfo {author}
  {\bibfnamefont {S.}~\bibnamefont {Fairhurst}}, \ and\ \bibinfo {author}
  {\bibfnamefont {D.}~\bibnamefont {Brown}},\ }\href@noop {} {\  (\bibinfo
  {year} {2017}{\natexlab{b}})}\BibitemShut {NoStop}%
\bibitem [{\citenamefont {Cannon}\ \emph {et~al.}(2012)\citenamefont {Cannon}
  \emph {et~al.}}]{Cannon:2011vi}%
  \BibitemOpen
  \bibfield  {author} {\bibinfo {author} {\bibfnamefont {K.}~\bibnamefont
  {Cannon}} \emph {et~al.},\ }\href {\doibase 10.1088/0004-637X/748/2/136}
  {\bibfield  {journal} {\bibinfo  {journal} {Astrophys. J.}\ }\textbf
  {\bibinfo {volume} {748}},\ \bibinfo {pages} {136} (\bibinfo {year}
  {2012})},\ \Eprint {http://arxiv.org/abs/1107.2665} {arXiv:1107.2665
  [astro-ph.IM]} \BibitemShut {NoStop}%
%%CITATION = ARXIV:1107.2665;%%
\bibitem [{\citenamefont {Privitera}\ \emph {et~al.}(2014)\citenamefont
  {Privitera}, \citenamefont {Mohapatra}, \citenamefont {Ajith}, \citenamefont
  {Cannon}, \citenamefont {Fotopoulos}, \citenamefont {Frei}, \citenamefont
  {Hanna}, \citenamefont {Weinstein},\ and\ \citenamefont
  {Whelan}}]{Privitera:2013xza}%
  \BibitemOpen
  \bibfield  {author} {\bibinfo {author} {\bibfnamefont {S.}~\bibnamefont
  {Privitera}}, \bibinfo {author} {\bibfnamefont {S.~R.~P.}\ \bibnamefont
  {Mohapatra}}, \bibinfo {author} {\bibfnamefont {P.}~\bibnamefont {Ajith}},
  \bibinfo {author} {\bibfnamefont {K.}~\bibnamefont {Cannon}}, \bibinfo
  {author} {\bibfnamefont {N.}~\bibnamefont {Fotopoulos}}, \bibinfo {author}
  {\bibfnamefont {M.~A.}\ \bibnamefont {Frei}}, \bibinfo {author}
  {\bibfnamefont {C.}~\bibnamefont {Hanna}}, \bibinfo {author} {\bibfnamefont
  {A.~J.}\ \bibnamefont {Weinstein}}, \ and\ \bibinfo {author} {\bibfnamefont
  {J.~T.}\ \bibnamefont {Whelan}},\ }\href {\doibase
  10.1103/PhysRevD.89.024003} {\bibfield  {journal} {\bibinfo  {journal} {Phys.
  Rev.}\ }\textbf {\bibinfo {volume} {D89}},\ \bibinfo {pages} {024003}
  (\bibinfo {year} {2014})},\ \Eprint {http://arxiv.org/abs/1310.5633}
  {arXiv:1310.5633 [gr-qc]} \BibitemShut {NoStop}%
%%CITATION = ARXIV:1310.5633;%%
\bibitem [{\citenamefont {Messick}\ \emph {et~al.}(2017)\citenamefont {Messick}
  \emph {et~al.}}]{Messick:2016aqy}%
  \BibitemOpen
  \bibfield  {author} {\bibinfo {author} {\bibfnamefont {C.}~\bibnamefont
  {Messick}} \emph {et~al.},\ }\href {\doibase 10.1103/PhysRevD.95.042001}
  {\bibfield  {journal} {\bibinfo  {journal} {Phys. Rev.}\ }\textbf {\bibinfo
  {volume} {D95}},\ \bibinfo {pages} {042001} (\bibinfo {year} {2017})},\
  \Eprint {http://arxiv.org/abs/1604.04324} {arXiv:1604.04324 [astro-ph.IM]}
  \BibitemShut {NoStop}%
%%CITATION = ARXIV:1604.04324;%%
\bibitem [{\citenamefont {Adams}\ \emph {et~al.}(2016)\citenamefont {Adams},
  \citenamefont {Buskulic}, \citenamefont {Germain}, \citenamefont {Guidi},
  \citenamefont {Marion}, \citenamefont {Montani}, \citenamefont {Mours},
  \citenamefont {Piergiovanni},\ and\ \citenamefont {Wang}}]{Adams:2015ulm}%
  \BibitemOpen
  \bibfield  {author} {\bibinfo {author} {\bibfnamefont {T.}~\bibnamefont
  {Adams}}, \bibinfo {author} {\bibfnamefont {D.}~\bibnamefont {Buskulic}},
  \bibinfo {author} {\bibfnamefont {V.}~\bibnamefont {Germain}}, \bibinfo
  {author} {\bibfnamefont {G.~M.}\ \bibnamefont {Guidi}}, \bibinfo {author}
  {\bibfnamefont {F.}~\bibnamefont {Marion}}, \bibinfo {author} {\bibfnamefont
  {M.}~\bibnamefont {Montani}}, \bibinfo {author} {\bibfnamefont
  {B.}~\bibnamefont {Mours}}, \bibinfo {author} {\bibfnamefont
  {F.}~\bibnamefont {Piergiovanni}}, \ and\ \bibinfo {author} {\bibfnamefont
  {G.}~\bibnamefont {Wang}},\ }\href {\doibase 10.1088/0264-9381/33/17/175012}
  {\bibfield  {journal} {\bibinfo  {journal} {Class. Quant. Grav.}\ }\textbf
  {\bibinfo {volume} {33}},\ \bibinfo {pages} {175012} (\bibinfo {year}
  {2016})},\ \Eprint {http://arxiv.org/abs/1512.02864} {arXiv:1512.02864
  [gr-qc]} \BibitemShut {NoStop}%
%%CITATION = ARXIV:1512.02864;%%
\bibitem [{\citenamefont {Keppel}(2013{\natexlab{a}})}]{Keppel:2013yia}%
  \BibitemOpen
  \bibfield  {author} {\bibinfo {author} {\bibfnamefont {D.}~\bibnamefont
  {Keppel}},\ }\href {\doibase 10.1103/PhysRevD.87.124003} {\bibfield
  {journal} {\bibinfo  {journal} {Phys. Rev.}\ }\textbf {\bibinfo {volume}
  {D87}},\ \bibinfo {pages} {124003} (\bibinfo {year} {2013}{\natexlab{a}})},\
  \Eprint {http://arxiv.org/abs/1303.2005} {arXiv:1303.2005 [physics.data-an]}
  \BibitemShut {NoStop}%
%%CITATION = ARXIV:1303.2005;%%
\bibitem [{\citenamefont {Abbott}\ \emph
  {et~al.}(2016{\natexlab{d}})\citenamefont {Abbott} \emph
  {et~al.}}]{Abbott:2016ymx}%
  \BibitemOpen
  \bibfield  {author} {\bibinfo {author} {\bibfnamefont {B.~P.}\ \bibnamefont
  {Abbott}} \emph {et~al.} (\bibinfo {collaboration} {Virgo, LIGO
  Scientific}),\ }\href {\doibase 10.3847/2041-8205/832/2/L21} {\bibfield
  {journal} {\bibinfo  {journal} {Astrophys. J.}\ }\textbf {\bibinfo {volume}
  {832}},\ \bibinfo {pages} {L21} (\bibinfo {year} {2016}{\natexlab{d}})},\
  \Eprint {http://arxiv.org/abs/1607.07456} {arXiv:1607.07456 [astro-ph.HE]}
  \BibitemShut {NoStop}%
%%CITATION = ARXIV:1607.07456;%%
\bibitem [{\citenamefont {Abbott}\ \emph
  {et~al.}(2017{\natexlab{a}})\citenamefont {Abbott} \emph {et~al.}}]{O1IMBH}%
  \BibitemOpen
  \bibfield  {author} {\bibinfo {author} {\bibfnamefont {B.~P.}\ \bibnamefont
  {Abbott}} \emph {et~al.} (\bibinfo {collaboration} {LIGO Scientific
  Collaboration, Virgo collaboration}),\ }\href@noop {} {\  (\bibinfo {year}
  {2017}{\natexlab{a}})},\ \Eprint {http://arxiv.org/abs/1704.04628}
  {arXiv:1704.04628 [gr-qc]} \BibitemShut {NoStop}%
%%CITATION = ARXIV:1704.04628;%%
\bibitem [{\citenamefont {Biswas}\ \emph {et~al.}(2012)\citenamefont {Biswas}
  \emph {et~al.}}]{Biswas:2012ty}%
  \BibitemOpen
  \bibfield  {author} {\bibinfo {author} {\bibfnamefont {R.}~\bibnamefont
  {Biswas}} \emph {et~al.},\ }\href {\doibase 10.1103/PhysRevD.85.122009}
  {\bibfield  {journal} {\bibinfo  {journal} {Phys. Rev.}\ }\textbf {\bibinfo
  {volume} {D85}},\ \bibinfo {pages} {122009} (\bibinfo {year} {2012})},\
  \Eprint {http://arxiv.org/abs/1201.2964} {arXiv:1201.2964 [gr-qc]}
  \BibitemShut {NoStop}%
%%CITATION = ARXIV:1201.2964;%%
\bibitem [{\citenamefont {Keppel}(2013{\natexlab{b}})}]{KeppelPlacement}%
  \BibitemOpen
  \bibfield  {author} {\bibinfo {author} {\bibfnamefont {D.}~\bibnamefont
  {Keppel}},\ }\href@noop {} {\  (\bibinfo {year} {2013}{\natexlab{b}})},\
  \Eprint {http://arxiv.org/abs/1307.4158} {arXiv:1307.4158 [gr-qc]}
  \BibitemShut {NoStop}%
%%CITATION = ARXIV:1307.4158;%%
\bibitem [{\citenamefont {Pürrer}(2014)}]{Purrer:2014fza}%
  \BibitemOpen
  \bibfield  {author} {\bibinfo {author} {\bibfnamefont {M.}~\bibnamefont
  {Pürrer}},\ }\href {\doibase 10.1088/0264-9381/31/19/195010} {\bibfield
  {journal} {\bibinfo  {journal} {Class. Quant. Grav.}\ }\textbf {\bibinfo
  {volume} {31}},\ \bibinfo {pages} {195010} (\bibinfo {year} {2014})},\
  \Eprint {http://arxiv.org/abs/1402.4146} {arXiv:1402.4146 [gr-qc]}
  \BibitemShut {NoStop}%
%%CITATION = ARXIV:1402.4146;%%
\bibitem [{\citenamefont {Taracchini}\ \emph {et~al.}(2014)\citenamefont
  {Taracchini} \emph {et~al.}}]{Taracchini:2013rva}%
  \BibitemOpen
  \bibfield  {author} {\bibinfo {author} {\bibfnamefont {A.}~\bibnamefont
  {Taracchini}} \emph {et~al.},\ }\href {\doibase 10.1103/PhysRevD.89.061502}
  {\bibfield  {journal} {\bibinfo  {journal} {Phys. Rev.}\ }\textbf {\bibinfo
  {volume} {D89}},\ \bibinfo {pages} {061502} (\bibinfo {year} {2014})},\
  \Eprint {http://arxiv.org/abs/1311.2544} {arXiv:1311.2544 [gr-qc]}
  \BibitemShut {NoStop}%
%%CITATION = ARXIV:1311.2544;%%
\bibitem [{\citenamefont {Buonanno}\ \emph {et~al.}(2009)\citenamefont
  {Buonanno}, \citenamefont {Iyer}, \citenamefont {Ochsner}, \citenamefont
  {Pan},\ and\ \citenamefont {Sathyaprakash}}]{Buonanno:2009zt}%
  \BibitemOpen
  \bibfield  {author} {\bibinfo {author} {\bibfnamefont {A.}~\bibnamefont
  {Buonanno}}, \bibinfo {author} {\bibfnamefont {B.}~\bibnamefont {Iyer}},
  \bibinfo {author} {\bibfnamefont {E.}~\bibnamefont {Ochsner}}, \bibinfo
  {author} {\bibfnamefont {Y.}~\bibnamefont {Pan}}, \ and\ \bibinfo {author}
  {\bibfnamefont {B.~S.}\ \bibnamefont {Sathyaprakash}},\ }\href {\doibase
  10.1103/PhysRevD.80.084043} {\bibfield  {journal} {\bibinfo  {journal} {Phys.
  Rev.}\ }\textbf {\bibinfo {volume} {D80}},\ \bibinfo {pages} {084043}
  (\bibinfo {year} {2009})},\ \Eprint {http://arxiv.org/abs/0907.0700}
  {arXiv:0907.0700 [gr-qc]} \BibitemShut {NoStop}%
%%CITATION = ARXIV:0907.0700;%%
\bibitem [{\citenamefont {Arun}\ \emph {et~al.}(2009)\citenamefont {Arun},
  \citenamefont {Buonanno}, \citenamefont {Faye},\ and\ \citenamefont
  {Ochsner}}]{Arun:2008kb}%
  \BibitemOpen
  \bibfield  {author} {\bibinfo {author} {\bibfnamefont {K.~G.}\ \bibnamefont
  {Arun}}, \bibinfo {author} {\bibfnamefont {A.}~\bibnamefont {Buonanno}},
  \bibinfo {author} {\bibfnamefont {G.}~\bibnamefont {Faye}}, \ and\ \bibinfo
  {author} {\bibfnamefont {E.}~\bibnamefont {Ochsner}},\ }\href {\doibase
  10.1103/PhysRevD.79.104023, 10.1103/PhysRevD.84.049901} {\bibfield  {journal}
  {\bibinfo  {journal} {Phys. Rev.}\ }\textbf {\bibinfo {volume} {D79}},\
  \bibinfo {pages} {104023} (\bibinfo {year} {2009})},\ \bibinfo {note}
  {[Erratum: Phys. Rev.D84,049901(2011)]},\ \Eprint
  {http://arxiv.org/abs/0810.5336} {arXiv:0810.5336 [gr-qc]} \BibitemShut
  {NoStop}%
%%CITATION = ARXIV:0810.5336;%%
\bibitem [{\citenamefont {Boh\'e}\ \emph {et~al.}(2017)\citenamefont {Boh\'e},
  \citenamefont {Shao}, \citenamefont {Taracchini}, \citenamefont {Buonanno},
  \citenamefont {Babak}, \citenamefont {Harry}, \citenamefont {Hinder},
  \citenamefont {Ossokine}, \citenamefont {P\"urrer}, \citenamefont {Raymond},
  \citenamefont {Chu}, \citenamefont {Fong}, \citenamefont {Kumar},
  \citenamefont {Pfeiffer}, \citenamefont {Boyle}, \citenamefont {Hemberger},
  \citenamefont {Kidder}, \citenamefont {Lovelace}, \citenamefont {Scheel},\
  and\ \citenamefont {Szil\'agyi}}]{PhysRevD.95.044028}%
  \BibitemOpen
  \bibfield  {author} {\bibinfo {author} {\bibfnamefont {A.}~\bibnamefont
  {Boh\'e}}, \bibinfo {author} {\bibfnamefont {L.}~\bibnamefont {Shao}},
  \bibinfo {author} {\bibfnamefont {A.}~\bibnamefont {Taracchini}}, \bibinfo
  {author} {\bibfnamefont {A.}~\bibnamefont {Buonanno}}, \bibinfo {author}
  {\bibfnamefont {S.}~\bibnamefont {Babak}}, \bibinfo {author} {\bibfnamefont
  {I.~W.}\ \bibnamefont {Harry}}, \bibinfo {author} {\bibfnamefont
  {I.}~\bibnamefont {Hinder}}, \bibinfo {author} {\bibfnamefont
  {S.}~\bibnamefont {Ossokine}}, \bibinfo {author} {\bibfnamefont
  {M.}~\bibnamefont {P\"urrer}}, \bibinfo {author} {\bibfnamefont
  {V.}~\bibnamefont {Raymond}}, \bibinfo {author} {\bibfnamefont
  {T.}~\bibnamefont {Chu}}, \bibinfo {author} {\bibfnamefont {H.}~\bibnamefont
  {Fong}}, \bibinfo {author} {\bibfnamefont {P.}~\bibnamefont {Kumar}},
  \bibinfo {author} {\bibfnamefont {H.~P.}\ \bibnamefont {Pfeiffer}}, \bibinfo
  {author} {\bibfnamefont {M.}~\bibnamefont {Boyle}}, \bibinfo {author}
  {\bibfnamefont {D.~A.}\ \bibnamefont {Hemberger}}, \bibinfo {author}
  {\bibfnamefont {L.~E.}\ \bibnamefont {Kidder}}, \bibinfo {author}
  {\bibfnamefont {G.}~\bibnamefont {Lovelace}}, \bibinfo {author}
  {\bibfnamefont {M.~A.}\ \bibnamefont {Scheel}}, \ and\ \bibinfo {author}
  {\bibfnamefont {B.}~\bibnamefont {Szil\'agyi}},\ }\href {\doibase
  10.1103/PhysRevD.95.044028} {\bibfield  {journal} {\bibinfo  {journal} {Phys.
  Rev. D}\ }\textbf {\bibinfo {volume} {95}},\ \bibinfo {pages} {044028}
  (\bibinfo {year} {2017})}\BibitemShut {NoStop}%
\bibitem [{\citenamefont {Bohé}\ \emph {et~al.}(2013)\citenamefont {Bohé},
  \citenamefont {Marsat},\ and\ \citenamefont {Blanchet}}]{Bohe:2013cla}%
  \BibitemOpen
  \bibfield  {author} {\bibinfo {author} {\bibfnamefont {A.}~\bibnamefont
  {Bohé}}, \bibinfo {author} {\bibfnamefont {S.}~\bibnamefont {Marsat}}, \
  and\ \bibinfo {author} {\bibfnamefont {L.}~\bibnamefont {Blanchet}},\ }\href
  {\doibase 10.1088/0264-9381/30/13/135009} {\bibfield  {journal} {\bibinfo
  {journal} {Class. Quant. Grav.}\ }\textbf {\bibinfo {volume} {30}},\ \bibinfo
  {pages} {135009} (\bibinfo {year} {2013})},\ \Eprint
  {http://arxiv.org/abs/1303.7412} {arXiv:1303.7412 [gr-qc]} \BibitemShut
  {NoStop}%
%%CITATION = ARXIV:1303.7412;%%
\bibitem [{lal()}]{lal}%
  \BibitemOpen
  \href@noop {} {\enquote {\bibinfo {title} {The {LIGO} algorithms library},}\
  }\bibinfo {howpublished}
  {\url{https://www.lsc-group.phys.uwm.edu/daswg/projects/lalsuite.html}}\BibitemShut
  {NoStop}%
\bibitem [{\citenamefont {Miller}\ and\ \citenamefont
  {Miller}(2014)}]{Miller:2014aaa}%
  \BibitemOpen
  \bibfield  {author} {\bibinfo {author} {\bibfnamefont {M.~C.}\ \bibnamefont
  {Miller}}\ and\ \bibinfo {author} {\bibfnamefont {J.~M.}\ \bibnamefont
  {Miller}},\ }\href {\doibase 10.1016/j.physrep.2014.09.003} {\bibfield
  {journal} {\bibinfo  {journal} {Phys.Rept.}\ }\textbf {\bibinfo {volume}
  {548}},\ \bibinfo {pages} {1} (\bibinfo {year} {2014})},\ \Eprint
  {http://arxiv.org/abs/1408.4145} {arXiv:1408.4145 [astro-ph.HE]} \BibitemShut
  {NoStop}%
\bibitem [{\citenamefont {Nitz}(2015)}]{AlexNitzThesis}%
  \BibitemOpen
  \bibfield  {author} {\bibinfo {author} {\bibfnamefont {A.}~\bibnamefont
  {Nitz}},\ }\href {http://surface.syr.edu/etd/316/} {Ph.D. thesis},\ \bibinfo
  {school} {Syracuse University} (\bibinfo {year} {2015})\BibitemShut {NoStop}%
%%CITATION = INSPIRE-1409173;%%
\bibitem [{\citenamefont {Lorimer}(2008)}]{Lorimer:2008se}%
  \BibitemOpen
  \bibfield  {author} {\bibinfo {author} {\bibfnamefont {D.~R.}\ \bibnamefont
  {Lorimer}},\ }\href {\doibase 10.12942/lrr-2008-8} {\bibfield  {journal}
  {\bibinfo  {journal} {Living Rev. Rel.}\ }\textbf {\bibinfo {volume} {11}},\
  \bibinfo {pages} {8} (\bibinfo {year} {2008})},\ \Eprint
  {http://arxiv.org/abs/0811.0762} {arXiv:0811.0762 [astro-ph]} \BibitemShut
  {NoStop}%
%%CITATION = ARXIV:0811.0762;%%
\bibitem [{\citenamefont {McClintock}\ \emph {et~al.}(2013)\citenamefont
  {McClintock}, \citenamefont {Narayan},\ and\ \citenamefont
  {Steiner}}]{McClintock:2013vwa}%
  \BibitemOpen
  \bibfield  {author} {\bibinfo {author} {\bibfnamefont {J.~E.}\ \bibnamefont
  {McClintock}}, \bibinfo {author} {\bibfnamefont {R.}~\bibnamefont {Narayan}},
  \ and\ \bibinfo {author} {\bibfnamefont {J.~F.}\ \bibnamefont {Steiner}},\
  }\href@noop {} {\  (\bibinfo {year} {2013})},\ \Eprint
  {http://arxiv.org/abs/1303.1583} {arXiv:1303.1583 [astro-ph.HE]} \BibitemShut
  {NoStop}%
\bibitem [{\citenamefont {Abbott}\ \emph
  {et~al.}(2016{\natexlab{e}})\citenamefont {Abbott} \emph
  {et~al.}}]{TheLIGOScientific:2016qqj}%
  \BibitemOpen
  \bibfield  {author} {\bibinfo {author} {\bibfnamefont {B.~P.}\ \bibnamefont
  {Abbott}} \emph {et~al.} (\bibinfo {collaboration} {Virgo, LIGO
  Scientific}),\ }\href {\doibase 10.1103/PhysRevD.93.122003} {\bibfield
  {journal} {\bibinfo  {journal} {Phys. Rev. D}\ }\textbf {\bibinfo {volume}
  {93}},\ \bibinfo {pages} {122003} (\bibinfo {year} {2016}{\natexlab{e}})},\
  \Eprint {http://arxiv.org/abs/1602.03839} {arXiv:1602.03839 [gr-qc]}
  \BibitemShut {NoStop}%
%%CITATION = ARXIV:1602.03839;%%
\bibitem [{\citenamefont {Klimenko}\ \emph {et~al.}(2016)\citenamefont
  {Klimenko} \emph {et~al.}}]{Klimenko:2015ypf}%
  \BibitemOpen
  \bibfield  {author} {\bibinfo {author} {\bibfnamefont {S.}~\bibnamefont
  {Klimenko}} \emph {et~al.},\ }\href {\doibase 10.1103/PhysRevD.93.042004}
  {\bibfield  {journal} {\bibinfo  {journal} {Phys. Rev.}\ }\textbf {\bibinfo
  {volume} {D93}},\ \bibinfo {pages} {042004} (\bibinfo {year} {2016})},\
  \Eprint {http://arxiv.org/abs/1511.05999} {arXiv:1511.05999 [gr-qc]}
  \BibitemShut {NoStop}%
%%CITATION = ARXIV:1511.05999;%%
\bibitem [{\citenamefont {Sathyaprakash}(1994)}]{PhysRevD.50.R7111}%
  \BibitemOpen
  \bibfield  {author} {\bibinfo {author} {\bibfnamefont {B.~S.}\ \bibnamefont
  {Sathyaprakash}},\ }\href {\doibase 10.1103/PhysRevD.50.R7111} {\bibfield
  {journal} {\bibinfo  {journal} {Phys. Rev. D}\ }\textbf {\bibinfo {volume}
  {50}},\ \bibinfo {pages} {R7111} (\bibinfo {year} {1994})}\BibitemShut
  {NoStop}%
\bibitem [{\citenamefont {Pürrer}(2016)}]{Purrer:2015tud}%
  \BibitemOpen
  \bibfield  {author} {\bibinfo {author} {\bibfnamefont {M.}~\bibnamefont
  {Pürrer}},\ }\href {\doibase 10.1103/PhysRevD.93.064041} {\bibfield
  {journal} {\bibinfo  {journal} {Phys. Rev.}\ }\textbf {\bibinfo {volume}
  {D93}},\ \bibinfo {pages} {064041} (\bibinfo {year} {2016})},\ \Eprint
  {http://arxiv.org/abs/1512.02248} {arXiv:1512.02248 [gr-qc]} \BibitemShut
  {NoStop}%
%%CITATION = ARXIV:1512.02248;%%
\bibitem [{\citenamefont {Gralla}\ \emph {et~al.}(2016)\citenamefont {Gralla},
  \citenamefont {Hughes},\ and\ \citenamefont {Warburton}}]{Gralla:2016qfw}%
  \BibitemOpen
  \bibfield  {author} {\bibinfo {author} {\bibfnamefont {S.~E.}\ \bibnamefont
  {Gralla}}, \bibinfo {author} {\bibfnamefont {S.~A.}\ \bibnamefont {Hughes}},
  \ and\ \bibinfo {author} {\bibfnamefont {N.}~\bibnamefont {Warburton}},\
  }\href {\doibase 10.1088/0264-9381/33/15/155002} {\bibfield  {journal}
  {\bibinfo  {journal} {Class. Quant. Grav.}\ }\textbf {\bibinfo {volume}
  {33}},\ \bibinfo {pages} {155002} (\bibinfo {year} {2016})},\ \Eprint
  {http://arxiv.org/abs/1603.01221} {arXiv:1603.01221 [gr-qc]} \BibitemShut
  {NoStop}%
%%CITATION = ARXIV:1603.01221;%%
\bibitem [{\citenamefont {Allen}(2005)}]{chi2veto}%
  \BibitemOpen
  \bibfield  {author} {\bibinfo {author} {\bibfnamefont {B.}~\bibnamefont
  {Allen}},\ }\href {\doibase 10.1103/PhysRevD.71.062001} {\bibfield  {journal}
  {\bibinfo  {journal} {Phys. Rev. D}\ }\textbf {\bibinfo {volume} {71}},\
  \bibinfo {pages} {062001} (\bibinfo {year} {2005})},\ \Eprint
  {http://arxiv.org/abs/gr-qc/0405045} {arXiv:gr-qc/0405045 [gr-qc]}
  \BibitemShut {NoStop}%
%%CITATION = GR-QC/0405045;%%
\bibitem [{\citenamefont {Babak}\ \emph {et~al.}(2013)\citenamefont {Babak},
  \citenamefont {Biswas}, \citenamefont {Brady}, \citenamefont {Brown} \emph
  {et~al.}}]{PhysRevD.87.024033}%
  \BibitemOpen
  \bibfield  {author} {\bibinfo {author} {\bibfnamefont {S.}~\bibnamefont
  {Babak}}, \bibinfo {author} {\bibfnamefont {R.}~\bibnamefont {Biswas}},
  \bibinfo {author} {\bibfnamefont {P.~R.}\ \bibnamefont {Brady}}, \bibinfo
  {author} {\bibfnamefont {D.~A.}\ \bibnamefont {Brown}},  \emph {et~al.},\
  }\href {\doibase 10.1103/PhysRevD.87.024033} {\bibfield  {journal} {\bibinfo
  {journal} {Phys. Rev. D}\ }\textbf {\bibinfo {volume} {87}},\ \bibinfo
  {pages} {024033} (\bibinfo {year} {2013})}\BibitemShut {NoStop}%
\bibitem [{\citenamefont {Abbott}\ \emph
  {et~al.}(2017{\natexlab{b}})\citenamefont {Abbott} \emph
  {et~al.}}]{PhysRevD.95.062003}%
  \BibitemOpen
  \bibfield  {author} {\bibinfo {author} {\bibfnamefont {B.~P.}\ \bibnamefont
  {Abbott}} \emph {et~al.} (\bibinfo {collaboration} {LIGO Scientific
  Collaboration}),\ }\href {\doibase 10.1103/PhysRevD.95.062003} {\bibfield
  {journal} {\bibinfo  {journal} {Phys. Rev. D}\ }\textbf {\bibinfo {volume}
  {95}},\ \bibinfo {pages} {062003} (\bibinfo {year}
  {2017}{\natexlab{b}})}\BibitemShut {NoStop}%
\bibitem [{\citenamefont {Dal~Canton}\ \emph
  {et~al.}(2014{\natexlab{b}})\citenamefont {Dal~Canton}, \citenamefont
  {Bhagwat}, \citenamefont {Dhurandhar},\ and\ \citenamefont
  {Lundgren}}]{0264-9381-31-1-015016}%
  \BibitemOpen
  \bibfield  {author} {\bibinfo {author} {\bibfnamefont {T.}~\bibnamefont
  {Dal~Canton}}, \bibinfo {author} {\bibfnamefont {S.}~\bibnamefont {Bhagwat}},
  \bibinfo {author} {\bibfnamefont {S.~V.}\ \bibnamefont {Dhurandhar}}, \ and\
  \bibinfo {author} {\bibfnamefont {A.}~\bibnamefont {Lundgren}},\ }\href
  {http://stacks.iop.org/0264-9381/31/i=1/a=015016} {\bibfield  {journal}
  {\bibinfo  {journal} {Classical and Quantum Gravity}\ }\textbf {\bibinfo
  {volume} {31}},\ \bibinfo {pages} {015016} (\bibinfo {year}
  {2014}{\natexlab{b}})},\ \Eprint {http://arxiv.org/abs/1304.0008}
  {arXiv:1304.0008 [gr-qc]} \BibitemShut {NoStop}%
\bibitem [{\citenamefont {Capano}\ \emph {et~al.}(2016)\citenamefont {Capano},
  \citenamefont {Harry}, \citenamefont {Privitera},\ and\ \citenamefont
  {Buonanno}}]{Capano:2016dsf}%
  \BibitemOpen
  \bibfield  {author} {\bibinfo {author} {\bibfnamefont {C.}~\bibnamefont
  {Capano}}, \bibinfo {author} {\bibfnamefont {I.}~\bibnamefont {Harry}},
  \bibinfo {author} {\bibfnamefont {S.}~\bibnamefont {Privitera}}, \ and\
  \bibinfo {author} {\bibfnamefont {A.}~\bibnamefont {Buonanno}},\ }\href
  {\doibase 10.1103/PhysRevD.93.124007} {\bibfield  {journal} {\bibinfo
  {journal} {Phys. Rev. D}\ }\textbf {\bibinfo {volume} {93}},\ \bibinfo
  {pages} {124007} (\bibinfo {year} {2016})},\ \Eprint
  {http://arxiv.org/abs/1602.03509} {arXiv:1602.03509 [gr-qc]} \BibitemShut
  {NoStop}%
%%CITATION = ARXIV:1602.03509;%%
\bibitem [{\citenamefont {Apostolatos}(1995)}]{PhysRevD.52.605}%
  \BibitemOpen
  \bibfield  {author} {\bibinfo {author} {\bibfnamefont {T.~A.}\ \bibnamefont
  {Apostolatos}},\ }\href {\doibase 10.1103/PhysRevD.52.605} {\bibfield
  {journal} {\bibinfo  {journal} {Phys. Rev. D}\ }\textbf {\bibinfo {volume}
  {52}},\ \bibinfo {pages} {605} (\bibinfo {year} {1995})}\BibitemShut
  {NoStop}%
\bibitem [{\citenamefont {Harry}\ \emph {et~al.}(2016)\citenamefont {Harry},
  \citenamefont {Privitera}, \citenamefont {Bohé},\ and\ \citenamefont
  {Buonanno}}]{Harry:2016ijz}%
  \BibitemOpen
  \bibfield  {author} {\bibinfo {author} {\bibfnamefont {I.}~\bibnamefont
  {Harry}}, \bibinfo {author} {\bibfnamefont {S.}~\bibnamefont {Privitera}},
  \bibinfo {author} {\bibfnamefont {A.}~\bibnamefont {Bohé}}, \ and\ \bibinfo
  {author} {\bibfnamefont {A.}~\bibnamefont {Buonanno}},\ }\href {\doibase
  10.1103/PhysRevD.94.024012} {\bibfield  {journal} {\bibinfo  {journal} {Phys.
  Rev.}\ }\textbf {\bibinfo {volume} {D94}},\ \bibinfo {pages} {024012}
  (\bibinfo {year} {2016})},\ \Eprint {http://arxiv.org/abs/1603.02444}
  {arXiv:1603.02444 [gr-qc]} \BibitemShut {NoStop}%
%%CITATION = ARXIV:1603.02444;%%
\bibitem [{\citenamefont {Miller}\ and\ \citenamefont
  {Colbert}(2004)}]{Miller:2003sc}%
  \BibitemOpen
  \bibfield  {author} {\bibinfo {author} {\bibfnamefont {M.~C.}\ \bibnamefont
  {Miller}}\ and\ \bibinfo {author} {\bibfnamefont {E.~J.~M.}\ \bibnamefont
  {Colbert}},\ }\href {\doibase 10.1142/S0218271804004426} {\bibfield
  {journal} {\bibinfo  {journal} {Int. J. Mod. Phys.}\ }\textbf {\bibinfo
  {volume} {D13}},\ \bibinfo {pages} {1} (\bibinfo {year} {2004})},\ \Eprint
  {http://arxiv.org/abs/astro-ph/0308402} {arXiv:astro-ph/0308402 [astro-ph]}
  \BibitemShut {NoStop}%
%%CITATION = ASTRO-PH/0308402;%%
\bibitem [{\citenamefont {Kalogera}(2000)}]{Kalogera:1999tq}%
  \BibitemOpen
  \bibfield  {author} {\bibinfo {author} {\bibfnamefont {V.}~\bibnamefont
  {Kalogera}},\ }\href {\doibase 10.1086/309400} {\bibfield  {journal}
  {\bibinfo  {journal} {Astrophys.J.}\ }\textbf {\bibinfo {volume} {541}},\
  \bibinfo {pages} {319} (\bibinfo {year} {2000})},\ \Eprint
  {http://arxiv.org/abs/astro-ph/9911417} {arXiv:astro-ph/9911417 [astro-ph]}
  \BibitemShut {NoStop}%
%%CITATION = ASTRO-PH/9911417;%%
\bibitem [{\citenamefont {Gerosa}\ \emph {et~al.}(2013)\citenamefont {Gerosa},
  \citenamefont {Kesden}, \citenamefont {Berti}, \citenamefont
  {O'Shaughnessy},\ and\ \citenamefont {Sperhake}}]{Gerosa:2013laa}%
  \BibitemOpen
  \bibfield  {author} {\bibinfo {author} {\bibfnamefont {D.}~\bibnamefont
  {Gerosa}}, \bibinfo {author} {\bibfnamefont {M.}~\bibnamefont {Kesden}},
  \bibinfo {author} {\bibfnamefont {E.}~\bibnamefont {Berti}}, \bibinfo
  {author} {\bibfnamefont {R.}~\bibnamefont {O'Shaughnessy}}, \ and\ \bibinfo
  {author} {\bibfnamefont {U.}~\bibnamefont {Sperhake}},\ }\href {\doibase
  10.1103/PhysRevD.87.104028} {\bibfield  {journal} {\bibinfo  {journal} {Phys.
  Rev.}\ }\textbf {\bibinfo {volume} {D87}},\ \bibinfo {pages} {104028}
  (\bibinfo {year} {2013})},\ \Eprint {http://arxiv.org/abs/1302.4442}
  {arXiv:1302.4442 [gr-qc]} \BibitemShut {NoStop}%
%%CITATION = ARXIV:1302.4442;%%
\bibitem [{\citenamefont {Farr}\ \emph {et~al.}(2011)\citenamefont {Farr},
  \citenamefont {Kremer}, \citenamefont {Lyutikov},\ and\ \citenamefont
  {Kalogera}}]{0004-637X-742-2-81}%
  \BibitemOpen
  \bibfield  {author} {\bibinfo {author} {\bibfnamefont {W.~M.}\ \bibnamefont
  {Farr}}, \bibinfo {author} {\bibfnamefont {K.}~\bibnamefont {Kremer}},
  \bibinfo {author} {\bibfnamefont {M.}~\bibnamefont {Lyutikov}}, \ and\
  \bibinfo {author} {\bibfnamefont {V.}~\bibnamefont {Kalogera}},\ }\href
  {http://stacks.iop.org/0004-637X/742/i=2/a=81} {\bibfield  {journal}
  {\bibinfo  {journal} {The Astrophysical Journal}\ }\textbf {\bibinfo {volume}
  {742}},\ \bibinfo {pages} {81} (\bibinfo {year} {2011})}\BibitemShut
  {NoStop}%
\bibitem [{\citenamefont {Rodriguez}\ \emph {et~al.}(2016)\citenamefont
  {Rodriguez}, \citenamefont {Zevin}, \citenamefont {Pankow}, \citenamefont
  {Kalogera},\ and\ \citenamefont {Rasio}}]{Rodriguez:2016vmx}%
  \BibitemOpen
  \bibfield  {author} {\bibinfo {author} {\bibfnamefont {C.~L.}\ \bibnamefont
  {Rodriguez}}, \bibinfo {author} {\bibfnamefont {M.}~\bibnamefont {Zevin}},
  \bibinfo {author} {\bibfnamefont {C.}~\bibnamefont {Pankow}}, \bibinfo
  {author} {\bibfnamefont {V.}~\bibnamefont {Kalogera}}, \ and\ \bibinfo
  {author} {\bibfnamefont {F.~A.}\ \bibnamefont {Rasio}},\ }\href {\doibase
  10.3847/2041-8205/832/1/L2} {\bibfield  {journal} {\bibinfo  {journal}
  {Astrophys. J.}\ }\textbf {\bibinfo {volume} {832}},\ \bibinfo {pages} {L2}
  (\bibinfo {year} {2016})},\ \Eprint {http://arxiv.org/abs/1609.05916}
  {arXiv:1609.05916 [astro-ph.HE]} \BibitemShut {NoStop}%
%%CITATION = ARXIV:1609.05916;%%
\bibitem [{\citenamefont {Brown}\ \emph {et~al.}(2012)\citenamefont {Brown},
  \citenamefont {Lundgren},\ and\ \citenamefont {O'Shaughnessy}}]{blo}%
  \BibitemOpen
  \bibfield  {author} {\bibinfo {author} {\bibfnamefont {D.~A.}\ \bibnamefont
  {Brown}}, \bibinfo {author} {\bibfnamefont {A.}~\bibnamefont {Lundgren}}, \
  and\ \bibinfo {author} {\bibfnamefont {R.}~\bibnamefont {O'Shaughnessy}},\
  }\href {\doibase 10.1103/PhysRevD.86.064020} {\bibfield  {journal} {\bibinfo
  {journal} {Phys. Rev. D}\ }\textbf {\bibinfo {volume} {86}},\ \bibinfo
  {pages} {064020} (\bibinfo {year} {2012})},\ \Eprint
  {http://arxiv.org/abs/1203.6060} {arXiv:1203.6060 [gr-qc]} \BibitemShut
  {NoStop}%
\bibitem [{\citenamefont {Dal~Canton}\ \emph {et~al.}(2015)\citenamefont
  {Dal~Canton}, \citenamefont {Lundgren},\ and\ \citenamefont
  {Nielsen}}]{PhysRevD.91.062010}%
  \BibitemOpen
  \bibfield  {author} {\bibinfo {author} {\bibfnamefont {T.}~\bibnamefont
  {Dal~Canton}}, \bibinfo {author} {\bibfnamefont {A.~P.}\ \bibnamefont
  {Lundgren}}, \ and\ \bibinfo {author} {\bibfnamefont {A.~B.}\ \bibnamefont
  {Nielsen}},\ }\href {\doibase 10.1103/PhysRevD.91.062010} {\bibfield
  {journal} {\bibinfo  {journal} {Phys. Rev. D}\ }\textbf {\bibinfo {volume}
  {91}},\ \bibinfo {pages} {062010} (\bibinfo {year} {2015})},\ \Eprint
  {http://arxiv.org/abs/1411.6815} {arXiv:1411.6815 [gr-qc]} \BibitemShut
  {NoStop}%
\bibitem [{\citenamefont {Indik}\ \emph {et~al.}(2016)\citenamefont {Indik},
  \citenamefont {Haris}, \citenamefont {Dal~Canton}, \citenamefont {Fehrmann},
  \citenamefont {Krishnan}, \citenamefont {Lundgren}, \citenamefont {Nielsen},\
  and\ \citenamefont {Pai}}]{Indik:2016qky}%
  \BibitemOpen
  \bibfield  {author} {\bibinfo {author} {\bibfnamefont {N.}~\bibnamefont
  {Indik}}, \bibinfo {author} {\bibfnamefont {K.}~\bibnamefont {Haris}},
  \bibinfo {author} {\bibfnamefont {T.}~\bibnamefont {Dal~Canton}}, \bibinfo
  {author} {\bibfnamefont {H.}~\bibnamefont {Fehrmann}}, \bibinfo {author}
  {\bibfnamefont {B.}~\bibnamefont {Krishnan}}, \bibinfo {author}
  {\bibfnamefont {A.}~\bibnamefont {Lundgren}}, \bibinfo {author}
  {\bibfnamefont {A.~B.}\ \bibnamefont {Nielsen}}, \ and\ \bibinfo {author}
  {\bibfnamefont {A.}~\bibnamefont {Pai}},\ }\href@noop {} {\  (\bibinfo {year}
  {2016})},\ \Eprint {http://arxiv.org/abs/1612.05173} {arXiv:1612.05173
  [gr-qc]} \BibitemShut {NoStop}%
%%CITATION = ARXIV:1612.05173;%%
\bibitem [{\citenamefont {Varma}\ \emph {et~al.}(2014)\citenamefont {Varma},
  \citenamefont {Ajith}, \citenamefont {Husa}, \citenamefont {Bustillo},
  \citenamefont {Hannam},\ and\ \citenamefont {Pürrer}}]{Varma:2014jxa}%
  \BibitemOpen
  \bibfield  {author} {\bibinfo {author} {\bibfnamefont {V.}~\bibnamefont
  {Varma}}, \bibinfo {author} {\bibfnamefont {P.}~\bibnamefont {Ajith}},
  \bibinfo {author} {\bibfnamefont {S.}~\bibnamefont {Husa}}, \bibinfo {author}
  {\bibfnamefont {J.~C.}\ \bibnamefont {Bustillo}}, \bibinfo {author}
  {\bibfnamefont {M.}~\bibnamefont {Hannam}}, \ and\ \bibinfo {author}
  {\bibfnamefont {M.}~\bibnamefont {Pürrer}},\ }\href {\doibase
  10.1103/PhysRevD.90.124004} {\bibfield  {journal} {\bibinfo  {journal} {Phys.
  Rev.}\ }\textbf {\bibinfo {volume} {D90}},\ \bibinfo {pages} {124004}
  (\bibinfo {year} {2014})},\ \Eprint {http://arxiv.org/abs/1409.2349}
  {arXiv:1409.2349 [gr-qc]} \BibitemShut {NoStop}%
%%CITATION = ARXIV:1409.2349;%%
\bibitem [{\citenamefont {Calderón~Bustillo}\ \emph
  {et~al.}(2016)\citenamefont {Calderón~Bustillo}, \citenamefont {Laguna},\
  and\ \citenamefont {Shoemaker}}]{Bustillo:2016gid}%
  \BibitemOpen
  \bibfield  {author} {\bibinfo {author} {\bibfnamefont {J.}~\bibnamefont
  {Calderón~Bustillo}}, \bibinfo {author} {\bibfnamefont {P.}~\bibnamefont
  {Laguna}}, \ and\ \bibinfo {author} {\bibfnamefont {D.}~\bibnamefont
  {Shoemaker}},\ }\href@noop {} {\  (\bibinfo {year} {2016})},\ \Eprint
  {http://arxiv.org/abs/1612.02340} {arXiv:1612.02340 [gr-qc]} \BibitemShut
  {NoStop}%
%%CITATION = ARXIV:1612.02340;%%
\bibitem [{\citenamefont {Capano}\ \emph {et~al.}(2014)\citenamefont {Capano},
  \citenamefont {Pan},\ and\ \citenamefont {Buonanno}}]{Capano:2013raa}%
  \BibitemOpen
  \bibfield  {author} {\bibinfo {author} {\bibfnamefont {C.}~\bibnamefont
  {Capano}}, \bibinfo {author} {\bibfnamefont {Y.}~\bibnamefont {Pan}}, \ and\
  \bibinfo {author} {\bibfnamefont {A.}~\bibnamefont {Buonanno}},\ }\href
  {\doibase 10.1103/PhysRevD.89.102003} {\bibfield  {journal} {\bibinfo
  {journal} {Phys. Rev.}\ }\textbf {\bibinfo {volume} {D89}},\ \bibinfo {pages}
  {102003} (\bibinfo {year} {2014})},\ \Eprint {http://arxiv.org/abs/1311.1286}
  {arXiv:1311.1286 [gr-qc]} \BibitemShut {NoStop}%
%%CITATION = ARXIV:1311.1286;%%
\bibitem [{\citenamefont {Moore}\ \emph {et~al.}(2016)\citenamefont {Moore},
  \citenamefont {Favata}, \citenamefont {Arun},\ and\ \citenamefont
  {Mishra}}]{Moore:2016qxz}%
  \BibitemOpen
  \bibfield  {author} {\bibinfo {author} {\bibfnamefont {B.}~\bibnamefont
  {Moore}}, \bibinfo {author} {\bibfnamefont {M.}~\bibnamefont {Favata}},
  \bibinfo {author} {\bibfnamefont {K.~G.}\ \bibnamefont {Arun}}, \ and\
  \bibinfo {author} {\bibfnamefont {C.~K.}\ \bibnamefont {Mishra}},\ }\href
  {\doibase 10.1103/PhysRevD.93.124061} {\bibfield  {journal} {\bibinfo
  {journal} {Phys. Rev.}\ }\textbf {\bibinfo {volume} {D93}},\ \bibinfo {pages}
  {124061} (\bibinfo {year} {2016})},\ \Eprint
  {http://arxiv.org/abs/1605.00304} {arXiv:1605.00304 [gr-qc]} \BibitemShut
  {NoStop}%
%%CITATION = ARXIV:1605.00304;%%
\bibitem [{\citenamefont {Tanay}\ \emph {et~al.}(2016)\citenamefont {Tanay},
  \citenamefont {Haney},\ and\ \citenamefont {Gopakumar}}]{Tanay:2016zog}%
  \BibitemOpen
  \bibfield  {author} {\bibinfo {author} {\bibfnamefont {S.}~\bibnamefont
  {Tanay}}, \bibinfo {author} {\bibfnamefont {M.}~\bibnamefont {Haney}}, \ and\
  \bibinfo {author} {\bibfnamefont {A.}~\bibnamefont {Gopakumar}},\ }\href
  {\doibase 10.1103/PhysRevD.93.064031} {\bibfield  {journal} {\bibinfo
  {journal} {Phys. Rev.}\ }\textbf {\bibinfo {volume} {D93}},\ \bibinfo {pages}
  {064031} (\bibinfo {year} {2016})},\ \Eprint
  {http://arxiv.org/abs/1602.03081} {arXiv:1602.03081 [gr-qc]} \BibitemShut
  {NoStop}%
%%CITATION = ARXIV:1602.03081;%%
\bibitem [{\citenamefont {Huerta}\ \emph {et~al.}(2017)\citenamefont {Huerta}
  \emph {et~al.}}]{Huerta:2016rwp}%
  \BibitemOpen
  \bibfield  {author} {\bibinfo {author} {\bibfnamefont {E.~A.}\ \bibnamefont
  {Huerta}} \emph {et~al.},\ }\href {\doibase 10.1103/PhysRevD.95.024038}
  {\bibfield  {journal} {\bibinfo  {journal} {Phys. Rev.}\ }\textbf {\bibinfo
  {volume} {D95}},\ \bibinfo {pages} {024038} (\bibinfo {year} {2017})},\
  \Eprint {http://arxiv.org/abs/1609.05933} {arXiv:1609.05933 [gr-qc]}
  \BibitemShut {NoStop}%
%%CITATION = ARXIV:1609.05933;%%
\bibitem [{\citenamefont {Hotokezaka}\ \emph {et~al.}(2011)\citenamefont
  {Hotokezaka}, \citenamefont {Kyutoku}, \citenamefont {Okawa}, \citenamefont
  {Shibata},\ and\ \citenamefont {Kiuchi}}]{PhysRevD.83.124008}%
  \BibitemOpen
  \bibfield  {author} {\bibinfo {author} {\bibfnamefont {K.}~\bibnamefont
  {Hotokezaka}}, \bibinfo {author} {\bibfnamefont {K.}~\bibnamefont {Kyutoku}},
  \bibinfo {author} {\bibfnamefont {H.}~\bibnamefont {Okawa}}, \bibinfo
  {author} {\bibfnamefont {M.}~\bibnamefont {Shibata}}, \ and\ \bibinfo
  {author} {\bibfnamefont {K.}~\bibnamefont {Kiuchi}},\ }\href {\doibase
  10.1103/PhysRevD.83.124008} {\bibfield  {journal} {\bibinfo  {journal} {Phys.
  Rev. D}\ }\textbf {\bibinfo {volume} {83}},\ \bibinfo {pages} {124008}
  (\bibinfo {year} {2011})}\BibitemShut {NoStop}%
\bibitem [{\citenamefont {Pannarale}\ \emph {et~al.}(2013)\citenamefont
  {Pannarale}, \citenamefont {Berti}, \citenamefont {Kyutoku},\ and\
  \citenamefont {Shibata}}]{PhysRevD.88.084011}%
  \BibitemOpen
  \bibfield  {author} {\bibinfo {author} {\bibfnamefont {F.}~\bibnamefont
  {Pannarale}}, \bibinfo {author} {\bibfnamefont {E.}~\bibnamefont {Berti}},
  \bibinfo {author} {\bibfnamefont {K.}~\bibnamefont {Kyutoku}}, \ and\
  \bibinfo {author} {\bibfnamefont {M.}~\bibnamefont {Shibata}},\ }\href
  {\doibase 10.1103/PhysRevD.88.084011} {\bibfield  {journal} {\bibinfo
  {journal} {Phys. Rev. D}\ }\textbf {\bibinfo {volume} {88}},\ \bibinfo
  {pages} {084011} (\bibinfo {year} {2013})}\BibitemShut {NoStop}%
\bibitem [{\citenamefont {Lackey}\ \emph {et~al.}(2014)\citenamefont {Lackey},
  \citenamefont {Kyutoku}, \citenamefont {Shibata}, \citenamefont {Brady},\
  and\ \citenamefont {Friedman}}]{PhysRevD.89.043009}%
  \BibitemOpen
  \bibfield  {author} {\bibinfo {author} {\bibfnamefont {B.~D.}\ \bibnamefont
  {Lackey}}, \bibinfo {author} {\bibfnamefont {K.}~\bibnamefont {Kyutoku}},
  \bibinfo {author} {\bibfnamefont {M.}~\bibnamefont {Shibata}}, \bibinfo
  {author} {\bibfnamefont {P.~R.}\ \bibnamefont {Brady}}, \ and\ \bibinfo
  {author} {\bibfnamefont {J.~L.}\ \bibnamefont {Friedman}},\ }\href {\doibase
  10.1103/PhysRevD.89.043009} {\bibfield  {journal} {\bibinfo  {journal} {Phys.
  Rev. D}\ }\textbf {\bibinfo {volume} {89}},\ \bibinfo {pages} {043009}
  (\bibinfo {year} {2014})}\BibitemShut {NoStop}%
\bibitem [{\citenamefont {Pannarale}\ \emph {et~al.}(2015)\citenamefont
  {Pannarale}, \citenamefont {Berti}, \citenamefont {Kyutoku}, \citenamefont
  {Lackey},\ and\ \citenamefont {Shibata}}]{Pannarale:2015jka}%
  \BibitemOpen
  \bibfield  {author} {\bibinfo {author} {\bibfnamefont {F.}~\bibnamefont
  {Pannarale}}, \bibinfo {author} {\bibfnamefont {E.}~\bibnamefont {Berti}},
  \bibinfo {author} {\bibfnamefont {K.}~\bibnamefont {Kyutoku}}, \bibinfo
  {author} {\bibfnamefont {B.~D.}\ \bibnamefont {Lackey}}, \ and\ \bibinfo
  {author} {\bibfnamefont {M.}~\bibnamefont {Shibata}},\ }\href {\doibase
  10.1103/PhysRevD.92.084050} {\bibfield  {journal} {\bibinfo  {journal} {Phys.
  Rev.}\ }\textbf {\bibinfo {volume} {D92}},\ \bibinfo {pages} {084050}
  (\bibinfo {year} {2015})},\ \Eprint {http://arxiv.org/abs/1509.00512}
  {arXiv:1509.00512 [gr-qc]} \BibitemShut {NoStop}%
%%CITATION = ARXIV:1509.00512;%%
\bibitem [{\citenamefont {van~der Walt}\ \emph {et~al.}(2011)\citenamefont
  {van~der Walt}, \citenamefont {Colbert},\ and\ \citenamefont
  {Varoquaux}}]{numpy}%
  \BibitemOpen
  \bibfield  {author} {\bibinfo {author} {\bibfnamefont {S.}~\bibnamefont
  {van~der Walt}}, \bibinfo {author} {\bibfnamefont {S.~C.}\ \bibnamefont
  {Colbert}}, \ and\ \bibinfo {author} {\bibfnamefont {G.}~\bibnamefont
  {Varoquaux}},\ }\href {\doibase http://dx.doi.org/10.1109/MCSE.2011.37}
  {\bibfield  {journal} {\bibinfo  {journal} {Computing in Science \&
  Engineering}\ }\textbf {\bibinfo {volume} {13}},\ \bibinfo {pages} {22}
  (\bibinfo {year} {2011})}\BibitemShut {NoStop}%
\bibitem [{\citenamefont {Hunter}(2007)}]{matplotlib}%
  \BibitemOpen
  \bibfield  {author} {\bibinfo {author} {\bibfnamefont {J.~D.}\ \bibnamefont
  {Hunter}},\ }\href {\doibase 10.1109/MCSE.2007.55} {\bibfield  {journal}
  {\bibinfo  {journal} {Computing In Science \& Engineering}\ }\textbf
  {\bibinfo {volume} {9}},\ \bibinfo {pages} {90} (\bibinfo {year}
  {2007})}\BibitemShut {NoStop}%
\end{thebibliography}%
\bibliographystyle{apsrev4-1}

\end{document}